\documentclass[aps,preprint]{revtex4}%
\usepackage{amsmath}
\usepackage{graphicx}
\usepackage{amsfonts}
\usepackage{amssymb}%
\begin{document}
\title{{\Large Preparation and control of a cavity-field state through atom-driven
field interaction: towards long-lived mesoscopic states}}
\author{C. J. Villas-B\^{o}as\thanks{E-mail: celsovb@df.ufscar.br}, F. R. de Paula, R.
M. Serra, and M. H. Y. Moussa}
\affiliation{Departamento de F\'{\i}sica, Universidade Federal de S\~{a}o Carlos, PO Box
676, S\~{a}o Carlos, 13565-905, SP, Brazil. }

\begin{abstract}
The preparation of mesoscopic states of the radiation and matter fields
through atom-field interactions has been achieved in recent years and employed
for a range of striking applications in quantum optics. Here we present a
technique for the preparation and control of a cavity mode which, besides
interacting with a two-level atom, is simultaneously submitted to linear and
parametric amplification processes. The role of the amplification-controlling
fields in the achievement of real mesoscopic states, is to produce
highly-squeezed field states and, consequently, to increase both: i) the
distance in phase space between the components of the prepared superpositions
and ii) the mean photon number of such superpositions. When submitting the
squeezed superposition states to the action of similarly squeezed reservoirs,
we demonstrate that under specific conditions the decoherence time of the
states becomes independent of both the distance in phase space between their
components and their mean photon number. An explanation is presented to
support this remarkable result, together with a discussion on the experimental
implementation of our proposal. We also show how to produce number states with
fidelities higher than those derived as circular states.


\textbf{Journal-ref: }Phys. Rev. A \textbf{68}, 053808 (2003)

\end{abstract}
\maketitle

\section{Introduction}

The successful manipulation of atom-field interactions in cavity quantum
electrodynamics (QED) and trapped ions is a great achievement of present-day
physics which has encouraged outstanding theoretical proposals and
experimental implementations. As high-$Q$ cavities \cite{Brune1} and ionic
traps \cite{Monroe2} have permitted the preparation of coherent-state
superpositions of the form $\left|  \Psi\right\rangle =$ $\left(  \left|
\alpha\operatorname{e}^{i\phi}\right\rangle +\left|  \alpha\operatorname{e}%
^{-i\phi}\right\rangle \right)  /\sqrt{2}$, with mean numbers of photon and
phonon quanta $\left|  \alpha\right|  ^{2}\approx10$, mesoscopic quantum
coherence has been investigated. In the cavity QED domain, the progressive
decoherence of mesoscopic superpositions involving radiation fields with
classically distinct phases was observed through atom-field interaction
\cite{Brune1} and the reversible decoherence of such a mesoscopic-field state
has been conjectured \cite{Raimond}. Moreover, the generation and detection of
Fock-states of the radiation field was demonstrated experimentally
\cite{Brattke} and the Rabi oscillation of circular Rydberg atoms in the
vacuum and in small coherent fields in a high-$Q$ cavity was measured
\cite{Brune2}, revealing the quantum nature of the radiation field
\cite{Knight1}.

Parallel to the achievements in cavity QED, the mastery of techniques to
manipulate electronic and motional states of trapped ions with classical
fields has enabled the control of fundamental quantum phenomena at a level
that seems to herald a new phase in technology. The operation of a two-bit
controlled-NOT quantum logic gate was demonstrated by storing the two quantum
bits in the internal and external degrees of freedom of a single trapped ion
\cite{Monroe1}. A ``Schr\"{o}dinger cat'' superposition of spatially-separated
coherent harmonic states was generated \cite{Monroe2}, as well as other
nonclassical states at the single atom level \cite{Meekhof}. The
reconstruction of the density matrices and the Wigner functions of various
quantum states of motion of a harmonically bound ion was also reported
\cite{Leibfried}.

Besides the atom-field interaction in cavity QED and trapped ions, the
preparation of \textit{reference} travelling-field states needed to measure
the properties of \textit{signal} travelling fields \cite{Barnett} has been
suggested based on optical linear \cite{Pegg} and nonlinear devices
\cite{D'Ariano}. The techniques developed over the last decades for the
process of parametric up- and down-conversion have enabled great advances in
the domain of travelling waves. The production of true entanglement by type-II
noncollinear phase-matching in parametric down-conversion was employed to
demonstrate a violation of Bell's inequality with two-photon fringe
visibilities in excess of 97\% \cite{Kwiat}. Three-photon
Greenberger-Horne-Zeilinger entanglement has also been observed \cite{Dik} and
it is worth stressing that the experimental implementations of teleportation
have been achieved with travelling wave techniques \cite{Teleportation}, as
these provide the facilities for preparation of the state to be teleported,
for the quantum channel and for the accomplishment of the required Bell-type
measurements. High-fidelity teleportation of superpositions \cite{Norton1} and
entanglements of running-wave field states \cite{Norton2} have also been presented.

As the techniques for generating nonclassical superposition states have been
improved, the attention turns to a major problem that must be overcome in the
contexts of quantum communication \cite{Comm} and computation \cite{Comp}: the
decoherence of quantum states due to the inevitable coupling of the quantum
systems to their environment \cite{Zurek,Caldeira,POA} and even due to
fluctuation in the interaction parameters required to prepare a coherent
superposition \cite{Bonifacio,Serra}. Schemes for inhibiting decoherence by
engineering the reservoir have been developed for trapped ions
\cite{Poyatos,Matos} and atomic two-level systems \cite{Lutkenhaus,Agarwal}.
Measurements of the decoherence of superposed motional states of a trapped ion
coupled to an engineered reservoir, where the coupling and the state of the
environment are controllable \cite{Myatt}, have also been reported. The
possibility of controlling the decoherence mechanism is crucial to the
preparation of the long-lived macroscopic superposition states and
entanglements of macroscopic objects required for the implementation of the
potential applications of quantum mechanics in communication and computation
\cite{JICirac}. Beyond the efforts being made to engineer mesoscopic
superpositions and entanglements with photon and phonon quanta, referred to
above, the possibility of engineering such mesoscopic states with massive
objects has been and is being pursued. Recently, correlations (on average) of
collective spin states of two macroscopic objects, each consisting of a
caesium gas sample with about $10^{12}$ atoms, was demonstrated
experimentally. In Ref. \cite{Zurek1} an experimental scenario designed to
reduce dramatically the decoherence rate of a quantum superposition of
Bose-Einstein condensates is outlined. This is also our concern in the present
work, focusing on the preparation of long-lived states of the radiation field
in cavity QED.

Methods for manipulating cavity-field states through atoms driven by external
fields \cite{Eberly} constitute an important means of attaining arbitrary
control of a quantum field. Although the time evolution of a field state under
linear and parametric amplifications has been a major concern in quantum
optics for generating squeezed states and investigating their properties
\cite{Scully,Dodonov}, of classical driving fields have barely been considered
for quantum states engineering purposes. Here we present a proposal for
achieving long-lived mesoscopic superposition states of the radiation field in
cavity QED which relies on two basic requirements: parametric amplification
and an engineered squeezed-vacuum reservoir for cavity-field states (we note
that the required engineered reservoir - resulting from the standard vacuum
for cavity modes plus additional interactions - must be an optimum
squeezed-vacuum reservoir). In addition, our technique can be employed to
prepare number states with fidelities higher than those generated as circular
states \cite{Janszki}.

Our proposal considers the dispersive interaction of a two-level atom with a
cavity field which is simultaneously under amplification processes. The
parametric amplification is employed to achieve a high degree of squeezing and
excitation of what we actually want to be a mesoscopic superposition state. We
show that the prepared squeezed-mesoscopic state, under the action of a
similarly squeezed reservoir, exhibits a decoherence time orders of magnitude
longer than those of non-squeezed cavity-field states subjected to the
influence of $i)$ a squeezed reservoir and $ii)$ a non-squeezed reservoir. In
fact, the computed decoherence time turns out to be independent of both the
average photon number and the distance in phase space between the centers of
the quasi-probability distribution of the individual states composing the
prepared superposition. The decoherence time depends only on the excitation of
the initial coherent state injected into the cavity previous to the
preparation of the squeezed superposition. This remarkable result follows when
the direction of squeezing of the superposition state is perpendicular to that
of the reservoir modes. Under this condition, the entanglement between the
prepared state and the modes of the reservoir is minimized and so the noise
injected from the reservoir into the prepared cavity mode is minimal, making
it a long-lived superposition state.

We finally stress that a scheme has been presented in Ref. \cite{Parametric}
for the implementation of the parametric amplification of an arbitrary
radiation-field state previously prepared in a high-$Q$ cavity. As squeezed
light is mainly supplied by nonlinear optical media as running waves (through
backward \cite{98} or forward \cite{99} four-wave mixing and parametric
down-conversion \cite{100}), standing squeezed fields in high-$Q$ cavities or
ion traps can be generated through atom-field interaction \cite{Meystre}.
Although considerable space has been devoted in the literature to the
squeezing process in the Jaynes-Cumming model, the issue of squeezing any
desired prepared cavity-field state $\left|  \Psi\right\rangle $, i.e., the
accomplishment of the operation $S(\zeta)\left|  \Psi\right\rangle $ in cavity
QED ($\zeta$ standing for a set of group parameters) has not been addressed.
Engineering such an operation was the subject of Ref. \cite{Parametric}; it is
achieved through the dispersive interactions of a three-level atom
simultaneously with a classical driving field and a cavity mode whose prepared
state we wish to squeeze. In short, the dispersive interaction of the cavity
mode with a driven atom produces the desired operation $S(\zeta)\left|
\Psi\right\rangle $. Since linear amplification is easily accomplished in
cavity QED \cite{LDavid,RMP}, the scheme in Ref. \cite{Parametric} contributes
crucially for the experimental feasibility of the present proposal for
preparation and control of long-lived cavity-field state through atom-driving
field interaction.

\section{Atom-driven field interaction}

The proposed configuration for engineering driven-cavity-field states,
depicted in Fig. 1, consists of a two-level Rydberg atom $A$ which crosses a
Ramsey-type arrangement, i.e., a high-$Q$ micromaser cavity $C$ located
between two Ramsey zones $R_{1}$ and $R_{2}$. After interacting with this
arrangement, the atom is counted by detection chambers $D_{2}$ and $D_{1}$
(for ionizing the excited $\left|  1\right\rangle $ and ground $\left|
0\right\rangle $ states, respectively), projecting the cavity-field in the
desired state. The transition of the two-level atom $A$ from excited to ground
state is far from resonant with the cavity mode frequency, allowing a
dispersive atom-field interaction to occur. In addition to the dispersive
interaction with the two-level atom, the cavity mode is simultaneously
submitted to linear and parametric amplifications (both represented in Fig. 1
by the source $S$) so that the Hamiltonian of our model (for $\hbar=1$) is
given by
\begin{equation}
H=\omega a^{\dagger}a+\frac{\omega_{0}}{2}\sigma_{z}+\chi a^{\dagger}%
a\sigma_{z}+\mathcal{H}_{amplification}\mathrm{{,}} \label{Eq1}%
\end{equation}
where $\sigma_{z}=|1\rangle\langle1|-|0\rangle\langle0|$, $a^{\dagger}$ and
$a$ are, respectively, the creation and annihilation operators for the cavity
mode of frequency $\omega$ which lies between the two atomic energy levels,
which are separated by $\omega_{0}$, such that the detuning $\delta
=|\omega-\omega_{0}|$ is large enough to enable only virtual transitions to
occur between the states $|0\rangle$ and $|1\rangle$. The atom-field coupling
parameter inside the cavity is $\chi=\Omega^{2}/\delta$ , where $\Omega$ is
the Rabi frequency. The expression for the atom-field dispersive interaction
on the right-hand side (rhs) of Eq. (\ref{Eq1}) is valid under the assumption
that $\Omega^{2}n\ll\delta^{2}+\gamma^{2}$, where $n$ is a characteristic
photon number and $\gamma$ is the spontaneous-emission rate \cite{Holland}. We
suppose, for simplicity, that the atom-field coupling is turned on (off)
suddenly at the instant the atom enters (leaves) the cavity region\textbf{,}
such that $\chi=0$ when the atom is outside the cavity.

We consider the atom, prepared at time $t_{0}$ by the Ramsey zone $R_{1}$ in a
$\left|  0\right\rangle $,$\left|  1\right\rangle $ superposition, to reach
$C$ at time $t_{1}$ and leave it at $t_{2}$. The linear and parametric pumping
are assumed to be turned on also at $t_{0}$ and turned off at a convenient
time $t\geq t_{2}$. Finally, the action of the classical amplification
mechanism on the cavity mode is described by the Hamiltonian
\begin{equation}
\mathcal{H}_{amplification}=\zeta(t)a^{\dagger^{2}}+\zeta^{\ast}(t)a^{^{2}%
}+\xi(t)a^{\dagger}+\xi^{\ast}(t)a\mathrm{{,}} \label{Eq1l}%
\end{equation}
where the time-dependent (TD) functions $\zeta(t)$ and $\xi(t)$ allow the
parametric and linear amplifications, respectively. It is well understood that
for specific values of these TD functions the eigenstates of the
Schr\"{o}dinger equation may squeeze the variance in one of the cavity modes'
two quadrature phases \cite{Scully,Dodonov,Walls,Salomon,Piza}.

The Schr\"{o}dinger state vector associated with Hamiltonian (\ref{Eq1}) can
be written using
\begin{equation}
|\Psi\left(  t\right)  \rangle=\operatorname*{e}\nolimits^{i\omega_{0}%
t/2}\left|  0\right\rangle \left|  \Phi_{0}\left(  t\right)  \right\rangle
+\operatorname*{e}\nolimits^{-i\omega_{0}t/2}\left|  1\right\rangle \left|
\Phi_{1}\left(  t\right)  \right\rangle \mathrm{{,}} \label{Eq2}%
\end{equation}
where $|\Phi_{\ell}\left(  t\right)  \rangle=\int\frac{d^{2}\alpha}{\pi
}\mathcal{A}_{\ell}\left(  \alpha,t\right)  |\alpha\rangle$, $\ell=0,1$, the
complex quantity $\alpha$ standing for the eigenvalues of $a$, and
$\mathcal{A}_{\ell}\left(  \alpha,t\right)  =\left\langle \alpha,\ell\left|
\Psi\left(  t\right)  \right.  \right\rangle $ are the expansion coefficients
for $|\Phi_{\ell}\left(  t\right)  \rangle$ in the coherent-state basis,
$\left\{  |\alpha\rangle\right\}  $. Using the orthogonality of the atomic
states and Eqs. (\ref{Eq1}) and (\ref{Eq2}) we obtain the uncoupled TD
Schr\"{o}dinger equations:
\begin{equation}
i\frac{d}{dt}|\Phi_{\ell}\left(  t\right)  \rangle=\mathbf{H}_{\ell}%
|\Phi_{\ell}\left(  t\right)  \rangle\mathrm{{,}} \label{Eq3}%
\end{equation}%
\begin{equation}
\mathbf{H}_{\ell}=\omega_{\ell}(t)a^{\dagger}a+\zeta(t)a^{\dagger^{2}}%
+\zeta^{\ast}(t)a^{^{2}}+\xi(t)a^{\dagger}+\xi^{\ast}(t)a\mathrm{{,}}
\label{Eq4}%
\end{equation}
with $\omega_{\ell}(t)=\left[  \omega-\left(  -1\right)  ^{\ell}\chi
\Theta(t-t_{1})\Theta(t_{2}-t)\right]  $. Note that the problem has been
reduced to that of a cavity field, under parametric and linear pumping, whose
frequency $\omega$ is shifted by $-\chi$ ($+\chi$) when interacting with the
atomic state $0$ ($1$), during the time interval $\tau=t_{2}-t_{1}$.

Solving Eq. (\ref{Eq3}) we obtain, from an initial state of the cavity
mode\textbf{\ }at time $t_{i}$, $\left|  \Phi_{\ell}(t_{i})\right\rangle $,
the evolved state \textbf{\ }
\begin{equation}
\left|  \Phi_{\ell}(t)\right\rangle =\mathbb{U}_{\ell}(t,t_{i})\left|
\Phi_{\ell}(t_{i})\right\rangle \mathrm{{,}} \label{Eq4l}%
\end{equation}
which defines the evolution operator we are looking for. Evidently, the
evolution operators $\mathbb{U}_{\ell}(t_{1},t_{0})$ and $\mathbb{U}_{\ell
}(t,t_{2})$, giving the evolution of the state vector of the radiation field
while the atom is outside the cavity, do not depend on the state of the
two-level atom, the label $\ell$ being unnecessary. However, the operator
$\mathbb{U}_{\ell}(t_{2},t_{1})$, which gives the evolution of the
cavity-field state during its interaction with the atom, does depend on the
atomic state and differs from the operators $\mathbb{U}(t_{1},t_{0})$ and
$\mathbb{U}(t,t_{2})$ only by the shifted frequency $\omega_{\ell}(t)$.

\section{ Solving the Schr\"{o}dinger equation via time-dependent invariants}

The Hamiltonian in Eq. (\ref{Eq4}) has been investigated in the search for
squeezed states of the radiation field. Group-theory methods
\cite{Walls,Aryeh} and TD invariants \cite{Salomon} have been used in attempts
to solve this TD quadratic Hamiltonian, which may represent a charged particle
subjected to a harmonic motion, immersed in a TD uniform magnetic field, a
single mode photon field travelling through a squeezing medium or, as in the
present situation, a cavity mode with shifted frequency under linear and
parametric amplification. In the present work, we make use of the TD
invariants of Lewis and Riesenfeld \cite{Lewis} to solve the Schr\"{o}dinger
equation (\ref{Eq3}), following the reasoning in Ref. \cite{Salomon}: instead
of proposing an invariant associated with the Hamiltonian (\ref{Eq4}), we
first perform a unitary transformation on Eq. (\ref{Eq3}) in order to reduce
it to a form which already has a known associated invariant. Thus, under a
unitary transformation represented by the operator $S(\varepsilon_{\ell})$
($\varepsilon_{\ell}$ standing for a set of TD group parameters which may also
depend on the atomic state $\ell$), we obtain from Eq. (\ref{Eq3})
\begin{equation}
i\frac{d}{dt}\left|  \Phi_{\ell}^{S}(t)\right\rangle =\mathcal{H}_{\ell}%
^{S}\left|  \Phi_{\ell}^{S}\left(  t\right)  \right\rangle \mathrm{{,}}
\label{Eq5}%
\end{equation}
where the transformed Hamiltonian and wave vector are given by
\begin{subequations}
\begin{align}
\mathcal{H}_{\ell}^{S}  &  =S^{\dagger}(\varepsilon_{\ell})\mathbf{H}_{\ell
}S(\varepsilon_{\ell})+i\frac{dS^{\dagger}(\varepsilon_{\ell})}{dt}%
S(\varepsilon_{\ell})\mathrm{{,}}\label{Eq5la}\\
\left|  \Phi_{\ell}^{S}\left(  t\right)  \right\rangle  &  =S^{\dagger
}(\varepsilon_{\ell})|\Phi_{\ell}\left(  t\right)  \rangle\mathrm{{.}}
\label{Eq5lb}%
\end{align}

In what follows we employ two theorems to obtain the solution of the TD
Schr\"{o}dinger equation (\ref{Eq3}): a) a theorem expounded in \cite{Salomon}
asserts that if $I_{\ell}(t)$ is an invariant associated with $\mathbf{H}%
_{\ell}$ (i.e., $dI_{\ell}(t)/dt=\partial I_{\ell}/\partial t+i\left[
\mathcal{H}_{\ell},I_{\ell}(t)\right]  =0$), then the transformed operator
$I_{\ell}^{S}(t)=S^{\dagger}(\varepsilon_{\ell})I_{\ell}(t)S(\varepsilon
_{\ell})$ will be an invariant associated with $\mathcal{H}_{\ell}^{S}$; b) on
the other hand, from Lewis and Riesenfeld's well-known theorem \cite{Lewis},
it follows that a solution of the Schr\"{o}dinger equation is an eigenstate of
the Hermitian invariant $I_{\ell}(t)$ multiplied by a TD phase factor. It
follows from a) and b) that the solutions of Eq. (\ref{Eq3}) are given by
\end{subequations}
\begin{equation}
|\Phi_{\ell,m}\left(  t\right)  \rangle=S(\varepsilon_{\ell})\left|
\Phi_{\ell,m}^{S}\left(  t\right)  \right\rangle =S(\varepsilon_{\ell
})\operatorname*{e}\nolimits^{i\phi_{\ell,m}^{S}(t)}\left|  m,t\right\rangle
_{S}\ \mathrm{{,\quad}}m=0,1,2,...\mathrm{{,}} \label{Eq6}%
\end{equation}
where $\left|  m,t\right\rangle _{S}$ is the eigenstate of the invariant
\cite{Puri} and the Lewis and Riesenfeld phase \cite{Lewis} obeys
\begin{equation}
\phi_{\ell,m}^{S}(t)=\int_{t_{i}}^{t}dt^{\prime}{}_{S}\left\langle
m,t^{\prime}\right|  \left(  i\frac{\partial}{\partial t^{\prime}}%
-\mathcal{H}_{\ell}^{S}\right)  \left|  m,t^{\prime}\right\rangle
_{S}\mathrm{{.}} \label{Eq7}%
\end{equation}
It is straightforward to verify that under the unitary transformation carried
out by the operator $S(\varepsilon_{\ell})$ the TD phase is invariant:
$\phi_{\ell,m}^{S}(t)=\phi_{\ell,m}(t)$.

\subsection{The transformed Hamiltonian}

Next, we associate the unitary transformation with the squeeze operator
$S(\varepsilon_{\ell})=\exp\left[  \frac{1}{2}\left(  \varepsilon_{\ell
}a^{\dagger^{2}}-\varepsilon_{\ell}^{\ast}a^{2}\right)  \right]  $, where the
complex TD function $\varepsilon_{\ell}=r_{\ell}(t)\operatorname*{e}%
\nolimits^{i\varphi_{\ell}(t)}$ includes the squeeze parameters $r_{\ell}(t)$
and $\varphi_{\ell}(t)$. $r_{\ell}(t)$ is associated with a squeeze factor,
while $\varphi_{\ell}(t)$ defines the squeezing direction in phase space.
Moreover, the TD parameters for the parametric and linear amplification
processes are written as $\zeta(t)=\kappa(t)\operatorname*{e}\nolimits^{i\eta
(t)}$ and $\xi(t)=\varkappa(t)\operatorname*{e}\nolimits^{i\varpi(t)}$,
respectively. The squeeze parameters ($r_{\ell}(t)$, $\varphi_{\ell}(t)$), the
amplification amplitudes ($\kappa(t)$,$\varkappa(t)$) and frequencies
($\eta(t)$,$\varpi(t)$) are all real TD functions. With the above assumptions
and after a lengthy calculation, the transformed Hamiltonian becomes
\begin{equation}
\mathcal{H}_{\ell}^{S}=\Omega_{\ell}(t)a^{\dagger}a+\Lambda_{\ell
}(t)a^{\dagger}+\Lambda_{\ell}^{\ast}(t)a+\digamma_{\ell}(t)\mathrm{{,}}
\label{Eq8}%
\end{equation}
provided that its TD coefficients satisfy
\begin{subequations}
\begin{align}
\Omega_{\ell}(t)  &  =\omega_{\ell}(t)+2\kappa(t)\tanh r_{\ell}(t)\cos\left(
\eta(t)-\varphi_{\ell}(t)\right)  \mathrm{{,}}\label{Eq9a}\\
\Lambda_{\ell}(t)  &  =\xi(t)\cosh r_{\ell}(t)+\xi^{\ast}(t)\operatorname*{e}%
\nolimits^{i\varphi_{\ell}(t)}\sinh r_{\ell}(t)\mathrm{{,}}\label{Eq9b}\\
\digamma_{\ell}(t)\mathrm{{}}  &  =\kappa(t)\tanh r_{\ell}(t)\cos\left(
\eta(t)-\varphi_{\ell}(t)\right)  \mathrm{{,}} \label{Eq9c}%
\end{align}
while the squeeze parameters $r_{\ell}(t)$ and $\varphi_{\ell}(t)$ are
determined by solving the coupled differential equations
\end{subequations}
\begin{subequations}
\begin{align}
\overset{.}{r}_{\ell}(t)  &  =2\kappa(t)\sin\left(  \eta(t)-\varphi_{\ell
}(t)\right)  \mathrm{{,}}\label{Eq10a}\\
\overset{.}{\varphi}_{\ell}(t)  &  =-2\omega_{\ell}(t)-4\kappa(t)\coth\left(
2r_{\ell}(t)\right)  \cos\left(  \eta(t)-\varphi_{\ell}(t)\right)
\mathrm{{.}} \label{Eq10b}%
\end{align}
It is evident from these relations that the TD group parameters $\varepsilon
_{\ell}(t)$, defining the unitary operator $S(\varepsilon_{\ell})$, depend on
the atomic state $\ell$, as assumed from the beginning. We finally mention
that we have associated the unitary transformation with the squeeze operator
since the parametric amplification described by Hamiltonian (\ref{Eq1l})
actually squeeze the cavity-field state. In fact, the TD parameter $\zeta(t)$
allowing the parametric amplification in Eq. (\ref{Eq1l}) is connected to the
squeeze parameters ($r_{\ell}(t)$, $\varphi_{\ell}(t)$) as expressed by Eqs.
(\ref{Eq10a}) and (\ref{Eq10b}).

\subsection{The evolution operators}

With the Hamiltonian (\ref{Eq8}) at hand we return to the solution of the
Schr\"{o}dinger equation (\ref{Eq5}). \smallskip The application of the
invariant method leads to the wave vector \cite{Puri}
\end{subequations}
\begin{equation}
\left|  \Phi_{\ell,m}^{S}(t)\right\rangle =\operatorname*{e}\nolimits^{i\phi
_{\ell,m}(t)}D\left[  \theta_{\ell}(t)\right]  \left|  m\right\rangle
\ \mathrm{{,\quad}}m=0,1,2,..., \label{Eq11}%
\end{equation}
where $\left|  m\right\rangle $ is the number state and $D\left[  \theta
_{\ell}(t)\right]  =\exp\left[  \theta_{\ell}(t)a^{\dagger}-\theta_{\ell
}^{\ast}(t)a\right]  $ is the displacement operator, $\theta_{\ell}(t)$ being
a solution to the equation $i\overset{.}{\theta}_{\ell}(t)=\Omega_{\ell
}(t)\theta_{\ell}(t)+\Lambda_{\ell}(t)$, given by
\begin{equation}
\theta_{\ell}(t)=\operatorname*{e}\nolimits^{-i\beta_{\ell}(t)}\left(
\theta_{\ell}(t_{i})-i\int_{t_{i}}^{t}\Lambda_{\ell}(t^{\prime}%
)\operatorname*{e}\nolimits^{i\beta_{\ell}(t^{\prime})}dt^{\prime}\right)
\mathrm{{,}} \label{Eq11l}%
\end{equation}
with $\beta_{\ell}(t)=\int_{t_{i}}^{t}\Omega_{\ell}(t^{\prime})dt^{\prime}$.
We note that $\theta_{\ell}(t_{0})$ describes the initial cavity-field state
which will be assumed to be a coherent state $\left|  \alpha\right\rangle $,
the subscript $\ell$ being purely formal. From the substitution of Hamiltonian
(\ref{Eq8}) into the Lewis and Riesenfeld phase, defined in Eq. (\ref{Eq7}),
we obtain
\begin{equation}
\phi_{\ell,m}(t)=-\int_{t_{i}}^{t}\left\{  m\Omega_{\ell}(t^{\prime})+\frac
{1}{2}\left[  \Lambda_{\ell}^{\ast}(t^{\prime})\theta_{\ell}(t^{\prime
})+\Lambda_{\ell}(t^{\prime})\theta_{\ell}^{\ast}(t^{\prime})\right]
+\digamma_{\ell}(t^{\prime})\right\}  dt^{\prime}\mathrm{{.}} \label{Eq12}%
\end{equation}

Therefore, the solutions of the Schr\"{o}dinger equation (\ref{Eq3}), which
form a complete set, can be written
\begin{equation}
\left|  \Phi_{\ell,m}(t)\right\rangle =S\left[  \varepsilon_{\ell}(t)\right]
\left|  \Phi_{\ell,m}^{S}(t)\right\rangle =U_{\ell}(t)\left|  m\right\rangle
\mathrm{{,}} \label{Eq13}%
\end{equation}
where
\begin{equation}
U_{\ell}(t)=\Upsilon_{\ell}(t)S\left[  \varepsilon_{\ell}(t)\right]  D\left[
\theta_{\ell}(t)\right]  R\left[  \Omega_{\ell}(t)\right]  \label{Eq13l}%
\end{equation}
is a unitary operator containing, in addition to the squeezed and the
displacement operators, a global phase factor
\begin{equation}
\Upsilon_{\ell}(t)=\exp\left\{  -i\int_{t_{i}}^{t}\left[  \operatorname{Re}%
\left[  \Lambda_{\ell}^{\ast}(t^{\prime})\theta_{\ell}(t^{\prime})\right]
+\digamma_{\ell}(t^{\prime})\right]  dt^{\prime}\right\}  \mathrm{{.}}
\label{Eq13ll}%
\end{equation}
The rotation operator in phase space, derived from the TD Lewis and Riesenfeld
phase factor, is given by
\begin{equation}
R\left[  \Omega_{\ell}(t)\right]  =\exp\left[  -ia^{\dagger}a\beta_{\ell
}(t)\right]  \mathrm{{,}} \label{Eq14}%
\end{equation}
Hence, for the solution of Schr\"{o}dinger equation \ref{Eq3}, we find
\begin{equation}
\left|  \Phi_{\ell}(t)\right\rangle =\sum\limits_{m=0}^{\infty}C_{m}\left|
\Phi_{\ell,m}(t)\right\rangle =U_{\ell}(t)\sum\limits_{m=0}^{\infty}%
C_{m}\left|  m\right\rangle =U_{\ell}(t)U_{\ell}^{\dagger}(t_{i})\left|
\Phi_{\ell}(t_{i})\right\rangle \mathrm{{,}} \label{Eq15}%
\end{equation}
which finally defines the evolution operators
\begin{equation}
\mathbb{U}_{\ell}(t,t_{i})=U_{\ell}(t)U_{\ell}^{\dagger}(t_{i}). \label{Eq15l}%
\end{equation}

We note that for the initial time $R\left[  \Omega_{\ell}(0)\right]  =R\left[
0\right]  =\mathbf{1}$, $D\left[  \theta_{\ell}(0)\right]  =D\left[
\alpha\right]  $, $S\left[  \varepsilon_{\ell}(0)\right]  =S\left[  0\right]
=\mathbf{1} $, and $\Upsilon_{\ell}(0)=1.$

\section{Evolution of the atom-field state}

Let us assume that the micromaser cavity is prepared at time $t_{0}$ in a
single-mode coherent state $|\alpha\rangle$\ by a monochromatic source, such
that with $m=0$ in Eq. (\ref{Eq13}) we have $\theta_{\ell}(t_{0})$ $=\alpha$.
Classical microwave fields are injected into the cavity and the amplitudes of
these fields can be adjusted by varying the injection time. As mentioned
above, the linear and parametric pumping are supposed to be turned on, also at
$t_{0}$, the same time the atom is prepared by the Ramsey zone $R_{1}$ in the
superposition state $c_{0}\left|  0\right\rangle +c_{1}\left|  1\right\rangle
$. The combined atom-field state at the initial time $t_{0}$ is, from Eq.
(\ref{Eq2})
\begin{equation}
|\Psi\left(  t_{0}\right)  \rangle=\left[  \operatorname*{e}\nolimits^{i\omega
_{0}t_{0}/2}c_{0}\left|  0\right\rangle +\operatorname*{e}\nolimits^{-i\omega
_{0}t_{0}/2}c_{1}\left|  1\right\rangle \right]  \left|  \alpha\right\rangle
\mathrm{{.}} \label{Eq16}%
\end{equation}
In fact, with $\mathcal{A}_{\ell}\left(  \beta,t_{0}\right)  =\left\langle
\beta,\ell\right|  \left(  c_{0}\left|  0\right\rangle +c_{1}\left|
1\right\rangle \right)  |\alpha\rangle$ it follows immediately that
$|\Phi_{\ell}\left(  t_{0}\right)  \rangle=\int\frac{d^{2}\beta}{\pi
}\mathcal{A}_{\ell}\left(  \beta,t_{0}\right)  |\beta\rangle=$ $c_{\ell
}\left|  \alpha\right\rangle $.

The evolution of the initial state $|\Psi\left(  t_{0}\right)  \rangle$ to the
time the atom reaches the cavity reads
\begin{equation}
|\Psi\left(  t_{1}\right)  \rangle=\mathbb{U}(t_{1},t_{0})|\Psi\left(
t_{0}\right)  \rangle\mathrm{{.}} \label{Eq17}%
\end{equation}
Evidently, the evolution operators $\mathbb{U}(t_{1},t_{0})$ and
$\mathbb{U}(t,t_{2})$, which govern the dynamics of the cavity-field state
while the atom is outside the cavity, do not depend on the state of the
two-level atom. On the other hand, during the time interval $\tau=t_{2}-t_{1}$
the atom spends inside the cavity the evolution of the entire system is
dictated by the operator $\mathbb{U}_{\ell}(t_{2},t_{1})=U_{\ell}%
(t_{2})U_{\ell}^{\dagger}(t_{1})$. This depends on the atomic state $\ell$ and
differs from the operators $\mathbb{U}(t_{1},t_{0})$ and $\mathbb{U}(t,t_{2})$
by the shifted frequency $\omega_{\ell}(t)$. Therefore, at the time the atom
leaves the cavity, the state of the atom-field system is given by
\begin{equation}
|\Psi\left(  t_{2}\right)  \rangle=\left[  \operatorname*{e}\nolimits^{i\omega
_{0}t_{2}/2}c_{0}\left|  0\right\rangle \mathbb{U}_{0}(t_{2},t_{1}%
)+\operatorname*{e}\nolimits^{-i\omega_{0}t_{2}/2}c_{1}\left|  1\right\rangle
\mathbb{U}_{1}(t_{2},t_{1})\right]  \mathbb{U}(t_{1},t_{0})\left|
\alpha\right\rangle \mathrm{{.}} \label{Eq18}%
\end{equation}
After crossing the cavity, the atom evolves freely from $t_{2}$ until the time
it reaches the second Ramsey zone, $R_{2}$. During this time interval, the
cavity mode continues to be pumped and the complete state of the system,
evolving under the operator $\mathbb{U}(t,t_{2})$, reads
\begin{equation}
|\Psi\left(  t\right)  \rangle=\mathbb{U}(t,t_{2})\left[  \operatorname*{e}%
\nolimits^{i\omega_{0}t/2}c_{0}\left|  0\right\rangle \mathbb{U}_{0}%
(t_{2},t_{1})+\operatorname*{e}\nolimits^{-i\omega_{0}t/2}c_{1}\left|
1\right\rangle \mathbb{U}_{1}(t_{2},t_{1})\right]  \mathbb{U}(t_{1}%
,t_{0})\left|  \alpha\right\rangle \mathrm{{.}} \label{Eq19}%
\end{equation}
Next, the atom crosses the Ramsey zone $R_{2}$, where a $\pi/2$ pulse is
applied, leading the atom-field system to the entangled state
\begin{align}
|\Psi\left(  t\right)  \rangle &  =\frac{1}{\sqrt{2}}\left\{  \left[
-\operatorname*{e}\nolimits^{i\omega_{0}t/2}c_{0}\mathsf{U}_{0}(t,t_{0}%
)+\operatorname*{e}\nolimits^{-i\omega_{0}t/2}c_{1}\mathsf{U}_{1}%
(t,t_{0})\right]  \left|  0\right\rangle \right. \nonumber\\
&  \left.  +\left[  \operatorname*{e}\nolimits^{i\omega_{0}t/2}c_{0}%
\mathsf{U}_{0}(t,t_{0})+\operatorname*{e}\nolimits^{-i\omega_{0}t/2}%
c_{1}\mathsf{U}_{1}(t,t_{0})\right]  \left|  1\right\rangle \right\}  \left|
\alpha\right\rangle \mathrm{{,}} \label{Eq20}%
\end{align}
where we have defined the operators
\begin{equation}
\mathsf{U}_{\ell}(t,t_{0})=\mathbb{U}(t,t_{2})\mathbb{U}_{\ell}(t_{2}%
,t_{1})\mathbb{U}(t_{1},t_{0})\mathrm{{.}} \label{Eq21}%
\end{equation}
Finally, measurement of the atomic state projects the cavity field into the
``Schr\"{o}dinger cat''-like state
\begin{equation}
|\Psi\left(  t\right)  \rangle=\mathcal{N}_{\pm}\left[  \pm\operatorname*{e}%
\nolimits^{i\omega_{0}t/2}c_{0}\mathsf{U}_{0}(t,t_{0})+\operatorname*{e}%
\nolimits^{-i\omega_{0}t/2}c_{1}\mathsf{U}_{1}(t,t_{0})\right]  \left|
\alpha\right\rangle \mathrm{{,}} \label{Eq22}%
\end{equation}
where the sign $+$ or $-$ occurs if the atom is detected in state $\left|
1\right\rangle $ or $\left|  0\right\rangle $, respectively, and
$\mathcal{N}_{\pm}$ refers to the normalization factors. From Eq. (\ref{Eq22})
it follows that, after measuring the atomic level used to generate the
superposition state of the radiation field, it is possible to control this
superposition by adjusting the TD amplification parameters $\kappa(t)$,
$\varkappa(t)$, $\eta(t)$, and $\varpi(t)$.

It is worth noting that expression (\ref{Eq22}) can be manipulated, employing
Eqs. (\ref{Eq21}), (\ref{Eq15l}), and Eq. (\ref{Eq13l}), to give the simple
form
\begin{align}
|\Psi\left(  t\right)  \rangle &  =\mathcal{N}_{\pm}\left[  \pm
\operatorname*{e}\nolimits^{i\omega_{0}t/2}c_{0}\Upsilon_{0}(t)S\left[
\varepsilon_{0}(t)\right]  \left|  \theta_{0}(t)\right\rangle
+\operatorname*{e}\nolimits^{-i\omega_{0}t/2}c_{1}\Upsilon_{1}(t)S\left[
\varepsilon_{1}(t)\right]  \left|  \theta_{1}(t)\right\rangle \right]
\nonumber\\
&  =\mathcal{N}_{\pm}\sum_{\ell=0}^{1}c_{\ell}(t)S\left[  \varepsilon_{\ell
}(t)\right]  D\left[  \theta_{\ell}(t)\right]  \left|  0\right\rangle
{\ }\nonumber\\
&  =\mathrm{{\ }}\mathcal{N}_{\pm}\sum_{\ell=0}^{1}c_{\ell}(t)S\left[
\varepsilon_{\ell}(t)\right]  \left|  \theta_{\ell}(t)\right\rangle
\mathrm{{,}} \label{Eq22N}%
\end{align}
where $c_{\ell}(t)=\pm\left(  \pm\right)  ^{\ell}\operatorname*{e}%
\nolimits^{\left(  -\right)  ^{\ell}i\omega_{0}t/2}c_{\ell}\Upsilon_{\ell}(t)$
and the amplitude of the coherent state $\left|  \theta_{\ell}(t)\right\rangle
$ follows from Eq. (\ref{Eq11l}).

\subsection{Passing $N$ atoms through the cavity}

Let us proceed to the construction of a cavity-field state by passing two or
more atoms through cavity $C$. It is easy to conclude from Eq. (\ref{Eq22})
that, after the passage of $N$ atoms through cavity $C$, each atom prepared in
the state $c_{0,k}\left|  0\right\rangle +c_{1,k}\left|  1\right\rangle $ by
$R_{1}$, $k=1,...,N$, we obtain the cavity-field state
\begin{equation}
|\Psi_{N}\left(  t\right)  \rangle=\mathcal{N}_{\pm}\prod_{k=1}^{N}\left[
\pm\operatorname*{e}\nolimits^{i\omega_{0}t/2}c_{0,k}\mathsf{U}_{0,k}%
(t_{f,k},t_{i,k})+\operatorname*{e}\nolimits^{-i\omega_{0}t/2}c_{1,k}%
\mathsf{U}_{1,k}(t_{f,k},t_{i,k})\right]  \left|  \alpha\right\rangle
\mathrm{{,}} \label{EqNew}%
\end{equation}
where $t_{i,k}$ stands for the time when the $k$-th atom is prepared by
$R_{1}$ and $t_{f,k}$ stands for the time when the $k$-th atom is detected,
assumed to be the same as $t_{i,k+1}$. Therefore, we obtain $\mathsf{U}%
_{\ell,k}(t_{f,k},t_{i,k})=\mathbb{U}(t_{f,k}=t_{i,k+1},t_{2,k})\mathbb{U}%
_{\ell}(t_{2,k},t_{1,k})\mathbb{U}(t_{1,k},t_{i,k}=t_{f,k-1})\mathrm{{.}}$
After some manipulation and using Eq. (\ref{Eq22N}), the state (\ref{EqNew})
can be simplified to the form
\begin{align}
|\Psi_{N}\left(  t\right)  \rangle &  =\mathcal{N}_{\pm}\sum_{\ell
_{1},...,\ell_{N}=1}^{2}\prod_{k=1}^{N}c_{\ell_{k}}(t)S\left[  \varepsilon
_{\ell_{1},...,\ell_{N}}(t)\right]  \left|  \theta_{\ell_{1},...,\ell_{N}%
}(t)\right\rangle \nonumber\\
&  =\mathcal{N}_{\pm}\sum_{k=1}^{2^{N}}C_{k}(t)S\left[  \Xi_{k}(t)\right]
\left|  \vartheta_{k}(t)\right\rangle , \label{EqNN}%
\end{align}
where we have replaced $\sum_{\ell_{1},...,\ell_{N}=1}^{2}$ by $\sum
_{k=1}^{2^{N}}$, i.e., $\Xi_{k}(t)\equiv r_{\ell_{1},...,\ell_{N}}%
(t)\exp\left(  i\varphi_{\ell_{1},...,\ell_{N}}(t)\right)  $ and
$\vartheta_{k}(t)\equiv\theta_{\ell_{1},...,\ell_{N}}(t)$.

\section{Analytical solutions of the characteristic equations (13a,13b)}

In this section we present some specific solutions of the characteristic
equations (\ref{Eq10a},\ref{Eq10b}), following a more detailed treatment in
\cite{Salomon}. We investigate the situation where the cavity mode $\left|
\alpha\right\rangle $ is resonant with the driving classical fields during the
time the atom is out of the cavity: from $t_{0}$ to $t_{1}$ and from $t_{2}$
to $t$. The parametric amplifier is assumed to operate in a degenerate mode in
which the \textit{signal} and the \textit{idler} frequencies coincide,
producing a single-mode driving field. In the resonant regime this single-mode
field has the same frequency $\omega$ as the cavity mode so that
$\eta(t)=-2\omega t$ \cite{Scully}. For the resonant linear amplifier it
follows that $\varpi(t)=-\omega t$. However, during the time that the atom is
inside the cavity, from $t_{1}$ to $t_{2}$, it pulls the mode frequency out of
resonance with the classical driving fields, establishing a dispersive regime
of the amplification process. Thus, in what follows we derive the solutions of
the coupled differential equations (\ref{Eq10a},\ref{Eq10b}) for the resonant
and dispersive regimes.

\subsection{Resonant amplification}

We start with the solution of the characteristic equations (\ref{Eq10a}%
,\ref{Eq10b}) for the resonant amplification that operates while the atom is
out of the cavity, from $t_{0}$ to $t_{1}$ and $t_{2}$ to $t$, when so that
$\omega_{\ell}(t)=\omega$, $\varpi(t)=-\omega t$ and $\eta(t)=-2\omega t$.
Defining $\omega=\overset{\cdot}{f}(t)$, $\varphi(t)=-2f(t)+g(t)$, and
$\eta(t)=-2f(t)+h(t)$, Eqs. (\ref{Eq10a},\ref{Eq10b}) become
\begin{subequations}
\begin{align}
\overset{\cdot}{r}(t)  &  =-2\kappa(t)\sin\left(  g(t)-h(t)\right)
\mathrm{{,}}\label{A1a}\\
\overset{\cdot}{g}(t)  &  =-4\kappa(t)\coth\left(  2r(t)\right)  \cos\left(
g(t)-h(t)\right)  \mathrm{{.}} \label{A1b}%
\end{align}
Assuming that $h(t)=h$ is a constant, the time dependence is eliminated from
Eqs. (\ref{A1a}), (\ref{A1b}) and we are left with the first-order
differential equation
\end{subequations}
\begin{equation}
\frac{dr}{dg}=\frac{1}{2}\tanh(2r)\tan(g-h)\mathrm{{.}} \label{A2}%
\end{equation}
After integrating Eq. (\ref{A2}) we obtain the constant of motion
\begin{equation}
\cos\left(  \varphi(t)-\eta(t)\right)  \sinh\left(  2r(t)\right)
=\mathcal{C}_{i}\mathrm{{,}} \label{A3}%
\end{equation}
with $\mathcal{C}_{i}$ depending on the initial values $r(t_{i})$,
$\varphi(t_{i})$, and $\eta(t_{i})$, where $i=0,2$. Thus, the solutions of
Eqs. (\ref{A1a}) and (\ref{A1b}), which apply under the condition $\cosh
^{2}\left(  2r(t)\right)  >1+\mathcal{C}_{i}^{2}$, are given by
\begin{subequations}
\begin{align}
\cosh\left(  2r(t)\right)   &  =\sqrt{1+\mathcal{C}_{i}^{2}}\cosh\left[
\cosh^{-1}\left(  \frac{\cosh2r(t_{i})}{\sqrt{1+\mathcal{C}_{i}^{2}}}\right)
\pm u(t,t_{i})\right]  \mathrm{{,}}\label{A4a}\\
\cos\left(  \varphi(t)-\eta(t)\right)   &  =\frac{\mathcal{C}_{i}}{\sqrt
{\cosh^{2}2r(t)-1}}\mathrm{{,}} \label{A4b}%
\end{align}
where
\end{subequations}
\begin{equation}
u(t,t_{i})=4\int_{t_{i}}^{t}\kappa(t)dt\mathrm{{.}} \label{A40}%
\end{equation}
Note that for $t_{i}=0$\ the $\cosh^{-1}$ term in Eq. (\ref{A4a}) is
null,\ and the signals $\pm$\ become irrelevant. However, for $t_{i}=t_{2}%
$\ we choose the sign that gives $r(t)\geq0$.

\subsection{Dispersive amplification}

Dispersive amplification occurs during the time the atom is inside the cavity,
shifting the mode frequency $\omega$ by $\chi=\Omega^{2}/\delta$, so that
$\omega_{\ell}=\omega\pm\chi$. Evidently, the amplification frequencies are
unaffected by the passage of the atom, so that $\eta(t)=-2\omega t$ and
$\varpi(t)=-\omega t$. Assuming that parameter $\kappa$ is time-independent
and defining $\varphi_{\ell}(t)-\eta(t)=f_{\ell}(t)$ and $\overset{\cdot}%
{\eta}(t)+$ $2\omega_{\ell}(t)=\overset{\cdot}{g}_{\ell}$, Eqs. (\ref{Eq10a}%
,\ref{Eq10b}) become
\begin{subequations}
\begin{align}
\overset{\cdot}{r}_{\ell}(t)  &  =2\kappa(t)\sin\left(  f_{\ell}(t)\right)
\mathrm{{,}}\label{A5a}\\
\overset{\cdot}{f}_{\ell}(t)  &  =-\overset{\cdot}{g}_{\ell}-4\kappa
(t)\coth\left(  2r_{\ell}(t)\right)  \cos\left(  f_{\ell}(t)\right)
\mathrm{{.}} \label{A5b}%
\end{align}
Since $\overset{\cdot}{g}_{\ell}=-(-1)^{\ell}2\chi$ is a constant, Eqs.
(\ref{A5a},\ref{A5b}) can be solved by quadrature, leading to a constant of
motion
\end{subequations}
\begin{equation}
\cosh\left(  2r_{\ell}(t)\right)  +\mathfrak{P}_{\ell}\cos\left(
\varphi_{\ell}(t)-\eta(t)\right)  \sinh\left(  2r_{\ell}(t)\right)
=\mathcal{C}_{1}\mathrm{{,}} \label{A6}%
\end{equation}
which now depends on the initial values $r(t_{1})$, $\varphi(t_{1})$, and
$\eta(t_{1})$. Despite the assumption that the atom-field coupling is turned
on (off) suddenly, these initial values must be computed from the solutions
for the resonant amplification regime at time $t_{1}$. With this procedure we
obtain the solutions for the resonant amplification ($r(t_{1})$,
$\varphi(t_{1})$) as the limit of those for the dispersive amplification
($r(t_{1})$, $\varphi(t_{1})$) as $\chi\rightarrow0$. The parameter
$\mathfrak{P}_{\ell}=-(-1)^{\ell}2\kappa/\chi$, defined for a constant
amplification amplitude $\kappa$, is an effective macroscopic coupling.
Therefore, for the dispersive regime, we find three different solutions,
depending on whether the coupling is strong ($\left|  \mathfrak{P}_{\ell
}\right|  >1$), weak ($\left|  \mathfrak{P}_{\ell}\right|  <1$), or critical
coupling ($\left|  \mathfrak{P}_{\ell}\right|  =1$).

a) With strong coupling ($\left|  \mathfrak{P}_{\ell}\right|  >1$), we have
the relations
\begin{subequations}
\begin{align}
\cosh\left(  2r_{\ell}(t)\right)   &  =\frac{1}{\mathfrak{P}_{\ell}^{2}%
-1}\left[  \frac{\operatorname*{e} {}^{h(t)}}{4}+\mathfrak{P}_{\ell}%
^{2}\left(  \mathcal{C}_{1}^{2}+\mathfrak{P}_{\ell}^{2}-1\right)
\operatorname*{e} {}^{-h(t)}-\mathcal{C}_{1}\right]  \mathrm{{,}}\label{A7a}\\
\cos\left(  \varphi_{\ell}(t)-\eta(t)\right)   &  =\frac{\mathcal{C}_{1}%
-\cosh\left(  2r_{\ell}(t)\right)  }{\mathfrak{P}_{\ell}\sinh\left(  2r_{\ell
}(t)\right)  }\mathrm{{,}} \label{A7b}%
\end{align}
where%

\end{subequations}
\begin{equation}
h(t)=\mp\frac{\sqrt{\mathfrak{P}_{\ell}^{2}-1}}{\left|  \mathfrak{P}_{\ell
}\right|  }u(t,t_{1})+\ln\left[  2\left|  \mathfrak{P}_{\ell}\right|  \left(
\sqrt{\left(  \mathfrak{P}_{\ell}^{2}-1\right)  \left(  \mathcal{C}_{1}%
^{2}-1\right)  }+\mathcal{C}_{1}\left|  \mathfrak{P}_{\ell}\right|  \right)
\right]  \mathrm{{,}} \label{A70}%
\end{equation}
the sign being chosen so that $r(t)\geq0$. The function $u(t,t_{1})$ is
defined by Eq. (\ref{A40}).

b) For the weak coupling regime ($\left|  \mathfrak{P}_{\ell}\right|  <1$),
the TD squeeze parameters when $\mathcal{C}_{1}>\sqrt{1-\mathfrak{P}_{\ell
}^{2}}$ are given by Eq. (\ref{A7b}) and
\begin{align}
\cosh\left(  2r_{\ell}(t)\right)   &  =\frac{\mathcal{C}_{1}}{1-\mathfrak{P}%
_{\ell}^{2}}\left\{  1-\frac{\left|  \mathfrak{P}_{\ell}\right|
\sqrt{\mathcal{C}_{1}^{2}+\mathfrak{P}_{\ell}^{2}-1}}{\mathcal{C}_{1}}\right.
\times\nonumber\\
&  \left.  \sin\left[  \pm\frac{\sqrt{1-\mathfrak{P}_{\ell}^{2}}}{\left|
\mathfrak{P}_{\ell}\right|  }u(t,t_{1})+\arcsin\left(  \frac{\mathcal{C}%
_{1}\left|  \mathfrak{P}_{\ell}\right|  }{\sqrt{\mathcal{C}_{1}^{2}%
+\mathfrak{P}_{\ell}^{2}-1}}\right)  \right]  \right\}  \label{A8}%
\end{align}

c) Finally, for critical coupling ($\left|  \mathfrak{P}_{\ell}\right|  =1$),
the TD squeeze parameters are given by Eq. (\ref{A7b}) and the solution for
$r_{\ell}(t)$ follows from the equation
\begin{equation}
\cosh\left(  2r_{\ell}(t)\right)  =\frac{1}{2\mathcal{C}_{1}}\left[
1+\mathcal{C}_{1}^{2}+\left(  \sqrt{\mathcal{C}_{1}(2\cosh(2r(t_{1}%
))-\mathcal{C}_{1})-1}\mp\mathcal{C}_{1}u(t,t_{1})\right)  ^{2}\right]
\mathrm{{,}} \label{A9}%
\end{equation}
the parameter $\ell$ being redundant. Note that in Eqs. (\ref{A7a}) and
(\ref{A8}) the parameter $\ell$ is also unnecessary since the rhs of
Eq.(\ref{A7a}) is an even function of $\mathfrak{P}_{\ell}$. Therefore, the
squeezing factor $r(t)$ does not depend on the atomic state, in contrast to
the squeezing direction in phase space defined by $\varphi_{\ell}(t)$.

>From the above solutions for the resonant and dispersive amplifications it is
straightforward to obtain the behavior of the TD squeeze parameters from time
$t_{0}$, when the classical driving fields are turned on simultaneously to the
preparation of the atomic state by $R_{1}$, to any instant $t$ after the
atom-field interaction. The time $t$ may be chosen to be before, after or in
the course of the ionization detection of the atomic state:

i) From $t_{0}$ to $t_{1}$, the squeeze parameters follow from Eqs.
(\ref{A4a}) and (\ref{A4b}). As mentioned above, such equations apply under
the condition $\cosh^{2}\left(  2r(t)\right)  >1+\mathcal{C}_{i}^{2}$, which
is always satisfied for $\mathcal{C}_{i}=0$, a value following from the
initial conditions $r(t_{0})$, $\varphi(t_{0})$, and $\eta(t_{0})$. In fact,
for an initial coherent state injected into the cavity: $r(t_{0})=0$. Assuming
the parameter $\kappa$ to be time-independent, together with $\mathcal{C}%
_{0}=0$, Eqs. (\ref{A4a}) and (\ref{A4b}) lead to the simplified solutions
\begin{subequations}
\begin{align}
r(t)  &  =2\kappa t\mathrm{{,}}\label{A10a}\\
\varphi(t)  &  =-2\omega t+\pi/2\mathrm{{.}} \label{A10b}%
\end{align}

ii) From $t_{1}$ to $t_{2}$ we have three possible solutions for the squeeze
parameters, depending on the coupling strength $\left|  \mathfrak{P}_{\ell
}\right|  $. These solutions follow from the above-described expressions, in
Eqs. (\ref{A7a}), (\ref{A7b}), (\ref{A8}), and (\ref{A9}), given that the
constant of motion $\mathcal{C}_{1}=\cosh\left(  4\kappa t_{1}\right)  $,
computed from Eqs. (\ref{A6}), (\ref{A10a}), and (\ref{A10b}) with $t=t_{1}$.
It is straightforward to observe in these equations the well-known threshold
in the behavior of the TD squeeze factor $r(t)$,\ following from the quadratic
TD Hamiltonian (\ref{Eq4}) \cite{Salomon}: $r(t)$\ increases monotonically for
$\left|  \mathfrak{P}_{\ell}\right|  \geq1$, while for $\left|  \mathfrak{P}%
_{\ell}\right|  <1$\ it oscillates periodically. In the present paper we are
interested in the weak coupling regime, where the squeeze parameters follow
from Eqs. (\ref{A7b}) and (\ref{A8}). We note that for realistic physical
parameters we achieve higher squeezing factor even in this regime.

iii) From $t_{2}$ to $t$ the squeeze parameters are again derived from Eqs.
(\ref{A4a}) and (\ref{A4b}). The constant of motion is computed from the
initial conditions $r(t_{2})$, $\varphi(t_{2})$, and $\eta(t_{2})$, which
depend on the strong, critical or weak coupling regimes. For weak coupling,
$\left|  \mathfrak{P}_{\ell}\right|  <1$, in which we are interested, the
constant of motion in Eq. (\ref{A3}), derived from Eqs. (\ref{A7b}) and
(\ref{A8}) and depending on the atomic state, reads
\end{subequations}
\begin{equation}
\mathcal{C}_{2,\ell}=\frac{\mathcal{C}_{1}-\cosh(2r(t_{2}))}{\mathfrak{P}%
_{\ell}}\mathrm{{.}} \label{A11}%
\end{equation}

\section{Wigner functions and fluctuations of the quadratures}

Now we analyze the states (\ref{Eq22N}) and (\ref{EqNN}), projected into the
cavity after the detection of one or several atoms, respectively, and
especially control of these states through the amplification parameters. From
here on we assume that the atom is detected in excited state $\left|
1\right\rangle $, so that $|\Psi\left(  t\right)  \rangle=\mathcal{N}_{+}%
\sum_{\ell=0}^{1}c_{\ell}(t)S\left[  \varepsilon_{\ell}(t)\right]  \left|
\theta_{\ell}(t)\right\rangle $ for (\ref{Eq22N}) and $|\Psi\left(  t\right)
\rangle=\mathcal{N}_{\pm}\sum_{k=1}^{2^{N}}C_{k}(t)S\left[  \Xi_{k}(t)\right]
\left|  \vartheta_{k}(t)\right\rangle $ for (\ref{EqNN}). After computing the
density operator of these cavity-field states, $\rho(t)=|\Psi\left(  t\right)
\rangle\left\langle \Psi\left(  t\right)  \right|  $, which reflects all the
properties of a quantum system -- such as superpositions and decoherence (when
fluctuating parameters are in order) -- the symmetric ordered characteristic
function, defined as in \cite{Scully}, follows:
\begin{equation}
\mathfrak{C}_{S}(\gamma,\gamma^{\ast},t)=\mathrm{{Tr}}\left(  \rho
(t)\operatorname*{e}\nolimits^{\gamma a^{\dagger}-\gamma^{\ast}a}\right)
=\left\langle \Psi\left(  t\right)  \right|  \operatorname*{e}%
\nolimits^{\gamma a^{\dagger}-\gamma^{\ast}a}|\Psi\left(  t\right)
\rangle\mathrm{{.}} \label{W1}%
\end{equation}
\qquad From the characteristic function $\mathfrak{C}_{S}(\gamma,\gamma^{\ast
},t)$ we define the Wigner distribution function \cite{Scully}
\begin{equation}
W(\eta,\eta^{\ast},t)=\frac{1}{\pi^{2}}\int d^{2}\gamma\mathfrak{C}_{S}%
(\gamma,\gamma^{\ast},t)\operatorname*{e}\nolimits^{\gamma^{\ast}\eta
-\gamma\eta^{\ast}}\mathrm{{.}} \label{Eq23}%
\end{equation}
which will be employed here to represent the quantum properties of the
cavity-field state conveniently in a three-dimensional Re$(\eta)$,Im$(\eta
)$,$W$ space. The result of the lengthy and somewhat involved integration over
the entire complex plane is presented in Appendix A, only for the state
(\ref{Eq22N}).

Next we analyze the fluctuations of the quadratures of the cavity mode,
defined as the dimensionless position and momentum operators, $X_{1}=\left(
a+a^{\dagger}\right)  $ and $X_{2}=-i(a-a^{\dagger})$, respectively. The
dynamic fluctuations for these quadrature operators in the cavity-field state,
given by
\begin{equation}
\left\langle \Delta X_{j}^{2}\right\rangle =\left\langle X_{j}^{2}%
\right\rangle -\left\langle X_{j}\right\rangle ^{2}\mathrm{{,}}j=1,2
\label{Eq24}%
\end{equation}
are obtained by computing the variances,
\begin{subequations}
\begin{align}
\left\langle \Delta a^{2}\right\rangle  &  =\left\langle a^{2}\right\rangle
-\left\langle a\right\rangle ^{2}\mathrm{{,}}\label{Eq25a}\\
\left\langle \Delta\left(  a^{\dagger}\right)  ^{2}\right\rangle  &
=\left\langle \left(  a^{\dagger}\right)  ^{2}\right\rangle -\left\langle
\left(  a^{\dagger}\right)  \right\rangle ^{2}\mathrm{{,}}\label{Eq25b}\\
\left\langle \Delta a^{\dagger}a\right\rangle  &  =\left\langle a^{\dagger
}a\right\rangle -\left\langle a^{\dagger}\right\rangle \left\langle
a\right\rangle \mathrm{{.}} \label{Eq25c}%
\end{align}
We note that with the above definitions for $X_{1}$ and $X_{2}$ we get for the
coherent state the minimum uncertainty value $\left\langle \Delta X_{j}%
^{2}\right\rangle =1$. The expected values of the normal ordered operators
defined in Eqs. (\ref{Eq25a}), (\ref{Eq25b}), and (\ref{Eq25c}) may be
conveniently evaluated with the help of the normal ordered characteristic
function
\end{subequations}
\begin{equation}
\mathfrak{C}_{N}(\gamma,\gamma^{\ast},t)=\mathrm{{Tr}}\left(  \rho
(t)\operatorname*{e}\nolimits^{\gamma a^{\dagger}}\operatorname*{e}%
\nolimits^{-\gamma^{\ast}a}\right)  =\operatorname*{e}\nolimits^{\left|
\gamma\right|  ^{2}/2}\mathfrak{C}_{S}(\gamma,\gamma^{\ast},t)\mathrm{{.}}
\label{Eq26}%
\end{equation}
>From this equation, if we want the normally ordered moments, it is easy to
derive the expression
\begin{equation}
\left\langle \left(  a^{\dagger}\right)  ^{n}a^{m}\right\rangle =\frac
{\partial^{n}}{\partial\gamma^{n}}\frac{\partial^{m}}{\partial(-\gamma^{\ast
})^{m}}\left.  \mathfrak{C}_{N}(\gamma,\gamma^{\ast},t)\right|  _{\gamma
=\gamma^{\ast}=0}\mathrm{{,}} \label{Eq27}%
\end{equation}
which is suitable for computing the variances in Eqs. (\ref{Eq25a}),
(\ref{Eq25b}), (\ref{Eq25c}), and (\ref{Eq24}).

\section{Protocols for the preparation of ``Schr\"{o}dinger cat''-like states
and number states}

\subsection{``Schr\"{o}dinger cat''-like states}

To prepare a particular superposition state from (\ref{Eq22N}) we follow a
three-step protocol. 1) First, we adjust the amplitude $\kappa$ of the
parametric amplification and the atom-field interaction time $\tau=t_{2}%
-t_{1}$ in order to obtain a particular angle $\Theta$ $=\left|  \varphi
_{1}(t_{2})-\varphi_{2}(t_{2})\right|  $ defined by the squeezing directions
of the states composing the ``Schr\"{o}dinger cat''-like superposition. 2)
Next, the desired excitation of the prepared state can be achieved by
manipulating the excitation of the initial coherent state injected into the
cavity, the amplitude of the linear amplification (that of the parametric
amplification has been fixed in the first step), and the time interval of the
amplification process. 3) Finally, the amplitude of both states composing the
superposition can be adjusted through the probability amplitudes of the atomic
superposition state prepared in the first Ramsey zone.

In Figs. $2$(a,b and c) we present some superposition states of the cavity
mode generated with the above protocol.\ In all these figures we have
considered an atom prepared in $R_{1}$ in the superposition $\left(  \left|
0\right\rangle +\left|  1\right\rangle \right)  /\sqrt{2}$. We have also
disregarded the linear amplification process while the parametric
amplification is switched off at $t=t_{2}$ when the atom leaves the cavity. In
the captions of Figs. $2$(a,b, and c) we present the fluctuations for the
quadrature operators and the parameters $r(t_{2})$ and $\Theta$ used for the
preparation of the desired states. Figs. 2(a,b) indicate the possibility to
control the squeezing directions of the quasi-probability distribution of the
individual states composing the prepared ``Schr\"{o}dinger cat''-like
superposition. This control will be extremely useful for generating number
state as circular squeezed states as shown below.

It is worth noting that a number of exotic\textit{ }reference states have
being requested for measuring properties of chosen field states. In Ref.
\cite{Pegg} the reciprocal-binomial state is requested as a reference field
for measuring the phase distribution of a chosen field without having to
obtain sufficient information to reconstruct its complete state. An extension
of the proposal in Ref. \cite{Pegg} was present for the $Q$-function
measurement where a convenient choice of a reference state allows us to
measure dispersions of quadrature operators \cite{Baseia}. Therefore, we hope
that the control of the squeezing directions of the components of
superposition states achieved through our scheme could also be employed to
generate these useful reference states. We also mention that the state in Fig.
$2$(c), considered in the analysis in Section VIII, is crucial for achieving
long-lived mesoscopic superposition states of the radiation field in cavity QED.

Finally, we recall that the amplification processes could be considered, after
the atom-field interaction, for controlling the prepared cavity-field state.
Both amplification processes can furnish excitation to the cavity mode, while
the parametric one is able to increase the degree of squeezing.

\subsection{Number states as circular squeezed states}

>From the present scheme of atom-driven field interaction it is possible to
generate number states with higher fidelity than those generated as circular
states, i.e., a superposition of $M$ coherent states having the same modulus
and uniformly distributed around a circle in the phase space \cite{Janszki}.
To do this, we have to pass $N$ atoms through the cavity, obtaining the state
defined in Eq. (\ref{EqNN}), where $M=2^{N}$. Remembering that $\Xi
_{k}(t)=r_{\ell_{1},...,\ell_{N}}(t)\exp\left(  i\varphi_{\ell_{1}%
,...,\ell_{N}}(t)\right)  $ and that the squeezing factor $r(t)$ does not
depend on the atomic state, differently from the squeezing direction in phase
space $\varphi_{\ell}(t)$, we get $\Xi_{k}(t)=r(t)\exp\left(  i\varphi
_{k}(t)\right)  $, where we have defined $\varphi_{k}(t)\equiv\varphi
_{\ell_{1},...,\ell_{N}}(t)$, with $k$ running from $1$ to $2^{N}$. With these
considerations, we obtain from Eq. (\ref{EqNN}) the photon distribution
function
\begin{align}
\mathcal{P}_{n}(t)  &  =\left|  \left\langle n\right.  |\Psi_{N}\left(
t\right)  \rangle\right|  ^{2}=\left|  \mathcal{N}_{+}\sum_{k=1}^{2^{N}}%
C_{k}(t)\left\langle n\right|  S\left[  \Xi_{k}(t)\right]  \left|
\vartheta_{k}(t)\right\rangle \right|  ^{2}\nonumber\\
&  =\left|  \mathcal{N}_{+}\right|  ^{2}\sum_{k,m=1}^{2^{N}}\frac{\left[
\tanh r(t)\right]  ^{n}}{2^{n}n!\cosh r(t)}\operatorname*{e}{}^{i\left[
\mathbb{\varphi}_{m}(t)-\mathbb{\varphi}_{k}(t)\right]  n/2}\nonumber\\
&  \times\exp\left\{  -\frac{1}{2}\left(  \left|  \vartheta_{k}(t)\right|
^{2}+\left|  \vartheta_{m}(t)\right|  ^{2}\right)  +\frac{1}{2}\tanh
r(t)\left[  \left(  \vartheta_{k}^{\ast}(t)\right)  ^{2}\operatorname*{e}%
\nolimits^{i\varphi_{k}(t)}+\left(  \vartheta_{m}(t)\right)  ^{2}%
\operatorname*{e}\nolimits^{-i\varphi_{m}(t)}\right]  \right\} \nonumber\\
&  \times H_{n}^{\ast}\left(  \frac{\vartheta_{k}(t)\operatorname*{e}%
\nolimits^{-i\varphi_{k}(t)/2}}{\sqrt{2\cosh r(t)\sinh r(t)}}\right)
H_{n}\left(  \frac{\vartheta_{m}(t)\operatorname*{e}\nolimits^{-i\varphi
_{m}(t)/2}}{\sqrt{2\cosh r(t)\sinh r(t)}}\right)  \label{N1}%
\end{align}
where $H_{n}(x)$ is the $n$-th Hermite polynomial evaluated at $x$. We have
assumed that all $N$ atoms were prepared in the same state $\left(  \left|
0\right\rangle +\left|  1\right\rangle \right)  /\sqrt{2}$ and detected in
their excited states.

In order to get the superposition of squeezed coherent states centered around
the origin of the phase space (as required to generate the number state) we
have to switch off the linear amplification process to obtain, from Eq.
(\ref{Eq9b}), $\Lambda_{\ell}(t)=0$, leading to coherent states having equal
amplitudes $\vartheta_{k}(t)\equiv\operatorname*{e}\nolimits^{-i\beta
_{_{\ell_{1},...,\ell_{N}}}(t)}\theta_{_{\ell_{1},...,\ell_{N}}}(t_{i})$
$=\operatorname*{e}\nolimits^{-i\beta_{k}(t)}\alpha$. In addition, assuming
$\alpha$ is real, we have to adjust $\beta_{k}(t)$ to $(1-k)\pi/N$ and
$\varphi_{k}(t)=(k-1)2\pi/N$, to get a symmetric distribution of these states
around the center of the phase space. With these adjustments (which are
achieved through the interaction times between the $N$ atoms and the cavity
mode and also through the parametric amplification parameters $\kappa
(t)$,$\eta(t)$) the photon distribution function simplifies to
\begin{align}
\mathcal{P}_{n}(t)  &  =\left|  \mathcal{N}_{+}\right|  ^{2}\sum
_{k,m=1}^{2^{N}}\frac{\left[  \tanh r(t)\right]  ^{n}}{n!\cosh r(t)}%
\operatorname*{e} \nolimits^{-\left|  \alpha\right|  ^{2}(1-\tanh
r(t))}\nonumber\\
&  \times\left[  H_{n}\left(  \frac{\left|  \alpha\right|  }{\sqrt{2\cosh
r(t)\sinh r(t)}}\right)  \right]  ^{2}\sum_{k,m=1}^{2^{N}}\operatorname*{e}
\nolimits^{i\left[  \mathbb{\varphi}_{m}(t)-\mathbb{\varphi}_{k}(t)\right]
n/2}\mathrm{{.}} \label{N2}%
\end{align}

In Fig. 3a we present the Wigner distribution function of the state generated
from the passage of $N=2$ atoms through a cavity initially prepared in the
coherent state $\left|  \alpha\right|  =7.4$ with $r=0.99$. These choices of
the parameters $\left|  \alpha\right|  $ and $r$ are considered in order to
maximize the photon distribution function for $n=8$, attaining $\mathcal{P}%
_{n=8}(t)=0.95$ which is exactly the fidelity $\left|  \left\langle 8\right|
\left.  \Psi_{N=2}(t)\right\rangle \right|  ^{2}$ of the prepared state with
respect to the number state $\left|  8\right\rangle $. The value $0.95$ is
considerably larger than that computed without the parametric amplification
process, when a circular state is generated with maximum fidelity $0.56$ with
respect to the number state $\left|  8\right\rangle $. The fidelity $0.56$ is
computed from an initial coherent state $\left|  \alpha\right|  =2.83$. In
Fig. 3b we plot the Wigner function of the state prepared from the passage of
$N=3$ atoms through a cavity initially prepared in the coherent state $\left|
\alpha\right|  =8$ with $r=0.67$. Here we obtain the optimal value
$\mathcal{P}_{n=16}(t)=0.99$, to be compared with the fidelity $0.79$ computed
when the amplification process is switched off and a circular state is
prepared, from an initial coherent state $\left|  \alpha\right|  =3.95$.

Note that with the passage of $N$ atoms though the cavity a family of number
states is obtained: $\left|  n=q2^{N}\right\rangle $ with $q=1,2,...$.
However, we stress that the fidelity of the prepared state decreases as the
integer $q$ increases. In Table I we present the states $\left|
n=q2^{3}\right\rangle $ for some values of $q$, in order to compare the
maximized fidelities computed from our model ($\mathcal{F}_{q}=\left|
\left\langle n=q2^{3}\right|  \left.  \Psi_{N=3}(t)\right\rangle \right|
^{2}$) with those derived from the circular states technique ($\emph{F}$). We
do not present the values of $\left|  \alpha\right|  $ and $r$ \ used to
calculated the fidelities.

We thus conclude that the atom-driven field process is suitable for preparing
number states with higher fidelity than those generated as circular states.
\ Next we present another application of the states generated by the
atom-driven field interaction.

\section{Preparing long-lived mesoscopic superposition states}

Evidently, the squeezed superposition in Eq. (\ref{Eq22N}) was ideally
prepared. In a real engineering process the dissipative mechanisms of the
cavity and the two-level atom, despite of the fluctuations intrinsic to their
interaction, must be taken into account. The complex calculations involved in
the engineering of quantum states under realistic quantum dissipation and
fluctuation conditions can be computed through the phenomenological-operator
approach presented in Refs. \cite{POA,Serra}. However, in this paper we will
not take into account the action of the reservoir during the preparation of
the squeezed superposition (\ref{Eq22N}). As usual, to estimate the
decoherence time, we next consider that an ideally prepared state is submitted
to the action of a quantum reservoir described by a collection of harmonic
oscillators whose Hamiltonian is $\mathrm{H}_{R}=\sum_{k}\hbar\omega_{k}%
b_{k}^{\dagger}b$. In addition, we will be interested in the action of a
vacuum-squeezed reservoir at absolute zero; its initial density operator reads
$\rho_{R}=\prod_{k}S_{k}\left|  0_{k}\right\rangle \left\langle 0_{k}\right|
S_{k}^{\dagger}$, $S_{k}$ being the squeezed operator for the $k$-th bath
oscillator mode. We are assuming here that, somehow, it is possible to
describe all the mechanisms of dissipation of the cavity in terms of the
action of a vacuum-squeezed reservoir. Describing the interaction between the
reservoir and the system (the cavity mode modeled as $\mathrm{H}_{S}%
=\hbar\omega a^{\dagger}a$) as $\mathrm{V}=$ $\sum_{k}\hbar(\lambda
_{k}a^{\dagger}b_{k}+\lambda_{k}^{\ast}ab_{k}^{\dagger})$, \ characterized by
the strengths $\lambda_{k}$ , the decoherence time deduced from the
idempotency defect of the reduced density operator of the cavity field, as
suggested in \cite{Piza1}, is given by
\begin{equation}
\frac{\hbar^{2}}{2\tau^{2}}=-\left\langle \mathrm{H}\right\rangle _{S,R}%
^{2}+\left\langle \left\langle \mathrm{H}\right\rangle _{S}^{2}\right\rangle
_{R}+\left\langle \left\langle \mathrm{H}\right\rangle _{R}^{2}\right\rangle
_{S}-\left\langle \mathrm{H}^{2}\right\rangle _{S,R} \label{D1}%
\end{equation}
where the Hamiltonian comprehends a sum of three terms $\mathrm{H=H}%
_{S}+\mathrm{H}_{R}+\mathrm{V}$. The average $\left\langle \mathrm{H}%
\right\rangle _{S}$ ($\left\langle \mathrm{H}\right\rangle _{R}$) is taken
with respect to the density matrix of the system (reservoir), given by
$\rho_{S}=|\Psi\left(  t\right)  \rangle\left\langle \Psi\left(  t\right)
\right|  $ ($\rho_{R}$), where $|\Psi\left(  t\right)  \rangle$ is given by
Eq. (\ref{Eq22N}). From Eq. (\ref{D1}), the decoherence time of the
cavity-field state is given by
\begin{equation}
\mathbf{\tau}=\frac{\mathbf{\tau}_{R}}{2\left|  (2N+1)\left(  \left\langle
a^{\dagger}\right\rangle \left\langle a\right\rangle -\left\langle a^{\dagger
}a\right\rangle \right)  +2\operatorname{Re}\left[  M\left(  \left\langle
a^{\dagger}\right\rangle ^{2}-\left\langle \left(  a^{\dagger}\right)
^{2}\right\rangle \right)  \right]  -N\right|  }, \label{D2}%
\end{equation}
where $\mathbf{\tau}_{R}$ is the relaxation time defined by the cavity,
$N=\sinh^{2}(\widetilde{r})$, and $M=-\operatorname{e}^{i\widetilde{\varphi}%
}\sinh(2\widetilde{r})/2$, $\widetilde{r}$ and $\widetilde{\varphi}$ being the
squeeze parameters of the vacuum reservoir \cite{Scully}. Here the mean values
are computed from the prepared squeezed superposition (\ref{Eq22N}). Since the
excitation of the initial coherent state $\alpha$ and the squeeze parameters
($r(t_{2}),\varphi_{\ell}(t_{2})$) have been fixed by the engineering
protocol, we note that Eq. (\ref{D2}) depends only on the reservoir squeeze
parameters ($\widetilde{r}$,$\widetilde{\varphi}$). Considering the situation
where $\alpha$ is real and $\left\langle \alpha\right|  \left.  -\alpha
\right\rangle =\exp(-2\alpha^{2})\approx0$ (implying $\alpha\gtrapprox\sqrt
{2}$), besides the assumption that $\varphi_{1}(t_{2})=\varphi$ and
$\varphi_{2}(t_{2})=\varphi+2n\pi$ ($n$ integer) (implying $\Theta=\left|
\varphi_{1}(t_{2})-\varphi_{2}(t_{2})\right|  =2n\pi$, i.e.,\ the states
composing the superposition (\ref{Eq22N}) are squeezed in the same direction),
we obtain
\begin{align}
\mathbf{\tau}  &  =\mathbf{\tau}_{R}\left|  1+\cosh(2\widetilde{r})\left[
2\alpha^{2}\cos\varphi\sinh(2r)-\left(  1+2\alpha^{2}\right)  \cosh
(2r)\right]  \right. \nonumber\\
&  -\sinh(2\widetilde{r})\left[  (1+2\alpha^{2})\cos(\widetilde{\varphi
}-\varphi)\sinh(2r)-\alpha^{2}\left(  \cos(\widetilde{\varphi}-2\varphi
)+\cos\widetilde{\varphi}\right)  \cosh(2r)\right. \nonumber\\
&  +\left.  \left.  \alpha^{2}\left(  \cos(\widetilde{\varphi}-2\varphi
)-\cos\widetilde{\varphi}\right)  \right]  \right|  ^{-1} \label{D21}%
\end{align}
where $r=r(t_{2})$. The maximization of the decoherence time $\mathbf{\tau}$
in Eq. (\ref{D21}) with respect to the parameters ($\widetilde{r}$%
,$\widetilde{\varphi}$), leads to the results
\begin{subequations}
\begin{align}
\widetilde{r}_{A}  &  \mathbf{=}r+\ln(1+4\alpha^{2})/4,\quad\widetilde
{\varphi}_{A}=0,\label{D3a}\\
\widetilde{r}_{B}  &  =r-\ln(1+4\alpha^{2})/4,\quad\widetilde{\varphi}_{B}%
=\pi, \label{D3b}%
\end{align}
which follow when we take $\varphi(t_{2})=$ $(2m+1)\pi$ and $\varphi
(t_{2})=2m\pi$ ($m$ integer), respectively. When $\Theta\neq2n\pi$, the
maximum of $\mathbf{\tau}$ turns out to be smaller than that for $\Theta
=2n\pi$, given either by the pair ($\widetilde{r}_{A}$, $\widetilde{\varphi
}_{A}$) or ($\widetilde{r}_{B}$, $\widetilde{\varphi}_{B}$). Observe that the
directions of squeezing of both states composing the superposition
(\ref{Eq22N}), defined by the angles $\varphi_{1}(t_{2})$ and $\varphi
_{2}(t_{2})$ has to be perpendicular to the direction of squeezing of the
vacuum reservoir.

Next, we compute the ``distance'' in phase space between the centers of the
quasi-probability distribution of the individual states composing the prepared
superposition (\ref{Eq22N}). This distance is defined by the quadratures of
the cavity field $X=(a^{\dagger}+a)/2$ and $Y=(a-a^{\dagger})/2i$, as
\end{subequations}
\begin{equation}
D=\left[  \left(  \left\langle X\right\rangle _{2}-\left\langle X\right\rangle
_{1}\right)  ^{2}+\left(  \left\langle Y\right\rangle _{2}-\left\langle
Y\right\rangle _{1}\right)  ^{2}\right]  ^{1/2}, \label{D4}%
\end{equation}
the subscripts $1$,$2$ referring to the two states composing the
superposition. Taking $\varphi_{1}(t_{2})=$ $\varphi_{2}(t_{2})=(2m+1)\pi$ or
$2m\pi$, the distance becomes $D=\left\langle X\right\rangle _{2}-\left\langle
X\right\rangle _{1}=2\alpha\exp(r)$ or $2\alpha\exp(-r) $, respectively. We
will focus on the case $\varphi_{1}(t_{2})=\varphi_{2}(t_{2})=$ $(2m_{1}%
+1)\pi$, since it results in a large distance $D$ between the two states
composing what we actually want to be a mesoscopic superposition. The
decoherence time and the mean photon number of the prepared state, obtained
from the values ($\widetilde{r}_{A}$, $\widetilde{\varphi}_{A}$), with
$\exp(-2\alpha^{2})\approx$ $0$, read
\begin{align}
\mathbf{\tau}  &  \approx\mathbf{\tau}_{R}/\alpha,\label{D5}\\
\left\langle n\right\rangle  &  =\left\langle a^{\dagger}a\right\rangle
\approx\alpha^{2}\exp(2r)+\sinh^{2}r. \label{D6}%
\end{align}
Remarkably, with the approximation $\exp(-2\alpha^{2})\approx0$, the
decoherence time for the prepared cavity-field state when $\varphi_{1}%
(t_{2})=$ $\varphi_{2}(t_{2})=(2m_{2}+1)\pi$ --- \textit{under the action of a
vacuum reservoir squeezed in the direction }$\widetilde{\varphi}_{A}=0$ ---
turns out to be practically independent of the parameter $r$ and thus of its
intensity $\left\langle n\right\rangle $ and distance $D$. Therefore, the
decoherence time (\ref{D5}) becomes practically independent of the quantities
which define the macroscopic character of the cavity-field state. From the
result in Eqs. (\ref{D5}) and (\ref{D6}) we conclude that it is convenient to
start from a coherent state $\alpha$ as small as possible (within the limit
$\exp(-2\alpha^{2})\approx$ $0$) and to adjust the macroscopic coupling
parameter $\left|  \mathfrak{P}_{\ell}\right|  $ in order to obtain a large
squeeze factor and, consequently, a large intensity of the prepared state and
a large distance $D$, since we are actually interested in mesoscopic
superpositions. We stress that even considering the weak coupling regime
($\left|  \mathfrak{P}_{\ell}\right|  <1$) we obtain, from Eqs.(\ref{A4a}) and
(\ref{A8}), large squeeze parameters: considering $\left|  \emph{P}_{\ell
}\right|  =0.1$, $\alpha=\sqrt{2}$, and the experimental running time about
$\ 2\times10^{-4}s$, we get a superposition state where $r\approx2$ and
$\left\langle n\right\rangle \approx10^{2}$ photons.

The mechanism behind this result is the degree of entanglement between the
prepared state and the modes of the reservoir, which depends on the relative
direction of their squeezing, defined by the angles $\varphi_{1}(t_{2}%
)=$\textit{\ }$\varphi_{2}(t_{2})$ and $\widetilde{\varphi}_{A}$. A result
supporting this argument is presented in \cite{Knight} where it is shown that
the injection of two modes, squeezed in perpendicular directions, in a $50/50$
beam splitter does not generate an entangled state. A careful analysis of the
dependence of the relative direction of squeezing on the degree of
entanglement between a prepared state and its multimode reservoir will be
presented in \cite{CRM}. Despite the fact that the mechanism behind the
long-lived mesoscopic superpositions is mainly the perpendicular squeezing
directions between the prepared state and the reservoir modes, the magnitude
of the parameter $r$ plays a crucial role in the present scheme for producing
the mesoscopic superposition by increasing both its intensity $\left\langle
n\right\rangle $ and distance $D$ in phase space.

The values presented above for $\mathbf{\tau}$, $\left\langle n\right\rangle
$, and $D$ are to be compared with those when considering a non-squeezed
($NS$) cavity-field state ($\left\langle n\right\rangle _{NS}=\alpha^{2}$,
$D_{NS}=2\alpha$ ) under the influence of $i)$ a squeezed reservoir, resulting
in the decoherence time $\mathbf{\tau}_{i}\approx\mathbf{\tau}_{R}/\alpha$,
and $ii)$ a non-squeezed reservoir, such that $\mathbf{\tau}_{ii}%
\approx\mathbf{\tau}_{R}/2\alpha^{2}$. Note that in both cases $i)$ and $ii)$
we obtain the ratios $\left\langle n\right\rangle /\left\langle n\right\rangle
_{NS}\approx\exp(2r)$ and $D/D_{NS}\approx\exp(r)$. Therefore, despite the
exponential increase in the ratios of both excitation and distance, we still
get $\mathbf{\tau\approx\tau}_{i}$ when comparing our results with previous
schemes in the literature, where a squeezed reservoir is assumed for the
enhancement of the decoherence time \cite{Kim}; for non-squeezed cavity-field
states and reservoir, we obtain a still better result $\mathbf{\tau
\approx\alpha\tau}_{ii}$.

\section{Discussion and Conclusion}

We have presented a scheme for the preparation and control of a cavity-field
state through atom-driven field interaction. The Lewis and Riesenfield
time-dependent invariants \cite{Lewis} were employed to obtain the eigenstates
of the cavity mode dispersively interacting with a two-level atom and
simultaneously under linear and parametric amplification processes. Protocols
for preparing particular superposition states and the number state were
presented. While relying on the manipulation of the initial states of the
cavity mode and the two-level atom, considered in previous schemes
\cite{Brune3}, our protocol also employs the time-dependent parameters
involved in the amplification sources to achieve particular superposition
states and number states. We plotted some interesting ``Schr\"{o}dinger
cat''-like states and number states generated as circular squeezed states. We
demonstrated that the number states generated as circular squeezed states
exhibit higher fidelities than those generated as circular states.

We have shown how to prepare truly mesoscopic ``Schr\"{o}dinger cat''-like
states of the cavity field, actually squeezed superposition states, through
their coupling to likewise squeezed reservoirs. When assuming that the
squeezing direction of the cavity field is perpendicular to that of the
reservoir modes, we found that the decoherence time of the prepared
superposition state depends only on the initial coherent state of the cavity
field from which the squeezed superposition is generated. Therefore, the
decoherence time is independent of the average photon number and the distance
in phase space between the centers of the quasi-probability distribution of
the individual states composing the squeezed superposition. This result
follows from the degree of entanglement between the prepared state and the
modes of the reservoir, which depends on the relative direction of their
squeezing. When the squeezing direction of the prepared superposition and that
of the reservoir modes is perpendicular, the noise injected from the reservoir
into the prepared cavity mode is minimized. A detailed analysis of the
dependence of the relative direction of squeezing on the degree of
entanglement between a prepared state and its multimode reservoir will be
presented in \cite{CRM}.

The experimental implementation of the proposed scheme relies on the
possibility of engineering a squeezed reservoir as well as of parametrically
driving cavity-field radiation. We stress that a scheme to realize physically
a squeezed bath for cavity modes, via quantum-nondemolition-mediated feedback,
has already been presented in Ref. \cite{Vitali}. However, the feedback
process in \cite{Vitali} does not eliminate the standard nonsqueezed bath and,
as we have pointed out, our scheme requires an optimal squeezed-vacuum
reservoir. The subject of quantum-reservoir engineering has attracted some
attention, specially in the domain of trapped ions \cite{Poyatos}; more
specifically, a scheme has been presented for the engineering of
squeezed-bath-type interactions to protect a two-level system against
decoherence \cite{Lutkenhaus}.

We emphasize that a proposal to implement the parametric amplification of an
arbitrary radiation-field state previously prepared in a high-$Q$ cavity is
presented in Ref. \cite{Parametric}. As mentioned above, in this proposal the
nonlinear process is accomplished through the dispersive interactions of a
single three-level atom simultaneously with a classical driving field and a
previously prepared cavity mode whose state we wish to squeeze. Moreover,
regarding parametric amplification of cavity fields, a technique was recently
suggested, based on pulsed excitation of semiconductor layers (on the cavity
walls) by laser radiation \cite{Carugno}. It is worth mention that all the
treatment developed above in the context of cavity quantum electrodynamics,
for delaying the decoherence process of a squeezed superposition by coupling
it to a vacuum-squeezed reservoir, can also be implemented in ion traps. We
finally mention that the proposal presented here should motivate future
theoretical and experimental investigations.

\subsection{Appendix A}

In this appendix we present the Wigner function computed from Eq. (\ref{Eq23})
and the relation
\begin{equation}
\mathfrak{C}_{S}(\gamma,\gamma^{\ast},t)=\operatorname*{e}\nolimits^{\left|
\gamma\right|  ^{2}/2}\mathfrak{C}_{A}(\gamma,\gamma^{\ast}%
,t)=\operatorname*{e}\nolimits^{\left|  \gamma\right|  ^{2}/2}\left\langle
\Psi\left(  t\right)  \right|  \operatorname*{e}\nolimits^{-\gamma^{\ast}%
a}\operatorname*{e}\nolimits^{\gamma a^{\dagger}}|\Psi\left(  t\right)
\rangle\mathrm{{,}} \label{Ap1}%
\end{equation}
derived from the antinormal ordered characteristic function
\begin{equation}
\mathfrak{C}_{A}(\gamma,\gamma^{\ast},t)=\mathrm{{Tr}}\left(  \rho
(t)\operatorname*{e}\nolimits^{-\gamma^{\ast}a}\operatorname*{e}%
\nolimits^{\gamma a^{\dagger}}\right)  =\int\frac{d^{2}\beta}{\pi}\left|
\left\langle \beta\right|  \left.  \Psi\left(  t\right)  \right\rangle
\right|  ^{2}\operatorname*{e}\nolimits^{-\gamma^{\ast}\beta+\gamma
\beta^{\dagger}} \label{Ap2}%
\end{equation}
First we have to compute, from Eq. (\ref{Eq22N}), the final cavity-field state
$|\Psi\left(  t\right)  \rangle$ which, after a lengthy calculation, becomes
\begin{equation}
|\Psi\left(  t\right)  \rangle=|\Psi\left(  t\right)  \rangle=\mathcal{N}%
_{+}\left[  c_{1}\Gamma_{1}(t)S(\varepsilon_{1}(t))\left|  \theta
_{1}(t)\right\rangle +c_{2}\Gamma_{2}(t)S(\varepsilon_{2}(t))\left|
\theta_{2}(t)\right\rangle \right]  \label{Ap3}%
\end{equation}
where the TD function $\Gamma_{\ell}(t)$ is defined in terms of that given by
Eq. (\ref{Eq13ll}), as
\begin{equation}
\Gamma_{\ell}(t)=\Upsilon_{\ell}(t)\Upsilon_{\ell}(t_{2})\Upsilon_{\ell}%
(t_{1}){,} \label{Ap4}%
\end{equation}
and $\theta_{\ell}(t)$ is defined by Eq. (\ref{Eq11l}). From Eqs. (\ref{Ap2})
and (\ref{Ap3}) we obtain, with $i,j=1,2$, the result
\begin{align}
\mathfrak{C}_{A}(\gamma,\gamma^{\ast},t)  &  =\sum_{i,j=1}^{2}\frac{K_{ij}%
}{\sqrt{1-4b_{i}b_{j}^{\ast}}}\nonumber\\
&  \times\exp\left(  \frac{(a_{i}+\gamma)(a_{j}^{\ast}+\gamma^{\ast}%
)+b_{i}(a_{j}^{\ast}-\gamma^{\ast})^{2}+b_{j}^{\ast}(a_{i}+\gamma)^{2}%
}{1-4b_{i}b_{j}^{\ast}}\right)  , \label{Ap5}%
\end{align}
where the TD function $K_{ij}$ reads
\begin{align}
K_{ij}  &  =\left|  \mathcal{N}_{+}\right|  ^{2}c_{i}c_{j}^{\ast}\Gamma
_{i}(t)\Gamma_{j}^{\ast}(t)\operatorname{sech}(r(t))\nonumber\\
&  \times\exp\left[  -\frac{1}{2}\left(  \left|  \theta_{i}(t)\right|
^{2}+\left|  \theta_{j}(t)\right|  ^{2}\right)  +\frac{1}{2}\tanh r(t)\left(
\operatorname{e}^{i\varphi_{i}(t)}\left(  \theta_{i}^{\ast}(t)\right)
^{2}+\operatorname{e}^{-i\varphi_{j}(t)}\left(  \theta_{j}(t)\right)
^{2}\right)  \right]  \mathrm{{,}} \label{Ap6}%
\end{align}
and
\begin{align}
a_{i}  &  =\theta_{i}(t)\operatorname{sech}r(t)\mathrm{{,}}\label{Ap7}\\
b_{i}  &  =-\frac{1}{2}\tanh r(t)\operatorname{e}^{i\varphi_{i}(t)}%
\mathrm{{.}} \label{Ap8}%
\end{align}
Note that for the weak coupling regime, $r_{1}(t)=r_{2}(t)=r(t)$. Finally,
from Eq. (\ref{Eq23}) and the characteristic function in Eq. (\ref{Ap5}) we
obtain the Wigner function
\begin{equation}
W(\eta,\eta^{\ast},t)=\sum_{i,j=1}^{2}\frac{Aij}{\sqrt{B_{ij}^{2}%
-4C_{ij}D_{ij}}}\exp\left(  \frac{C_{ij}E_{ij}^{2}+D_{ij}F_{ij}^{2}%
+B_{ij}E_{ij}F_{ij}}{B_{ij}^{2}-4C_{ij}D_{ij}}\right)  \mathrm{{,}}
\label{Ap9}%
\end{equation}
where the TD functions $A_{ij}$, $B_{ij}$, $C_{ij}$, $D_{ij}$,.$E_{ij}$, and
$F_{ij}$ satisfy
\begin{align}
A_{ij}  &  =\frac{K_{ij}}{\sqrt{1-4b_{i}b_{j}^{\ast}}}\exp\left[  \frac
{a_{i}(t)a_{j}^{\ast}(t)+a_{i}^{2}(t)b_{j}^{\ast}(t)+\left(  a_{j}^{\ast
}(t)\right)  ^{2}b_{i}(t)}{1-4b_{i}(t)b_{j}^{\ast}(t)}\right] \label{Ap10}\\
B_{ij}  &  =-\frac{1}{2}+\frac{1}{1-4b_{i}(t)b_{j}^{\ast}(t)},\label{Ap11}\\
C_{ij}  &  =\frac{b_{i}(t)}{1-4b_{i}(t)b_{j}^{\ast}(t)},\label{Ap12}\\
D_{ij}  &  =\frac{b_{j}^{\ast}(t)}{1-4b_{i}(t)b_{j}^{\ast}(t)},\label{Ap13}\\
E_{ij}  &  =-\eta^{\ast}+\frac{2a_{i}(t)b_{j}^{\ast}(t)+a_{j}^{\ast}%
(t)}{1-4b_{i}(t)b_{j}^{\ast}(t)},\label{Ap14}\\
F_{ij}  &  =\eta-\frac{2a_{j}^{\ast}(t)b_{i}(t)+a_{i}(t)}{1-4b_{i}%
(t)b_{j}^{\ast}(t)}. \label{Ap15}%
\end{align}

\begin{acknowledgments}
We wish to express our thanks for the support of FAPESP (under contracts
\#99/11617-0, \#00/15084-5, and \#02/02633-6) and CNPq (Instituto do
Mil\^{e}nio de Informa\c{c}\~{a}o Qu\^{a}ntica), Brazilian research funding
agencies, and to Profs. S. S. Mizrahi, B. Baseia, N. G. de Almeida and R.
Napolitano for helpful discussions.
\end{acknowledgments}

\textbf{Figure Captions}

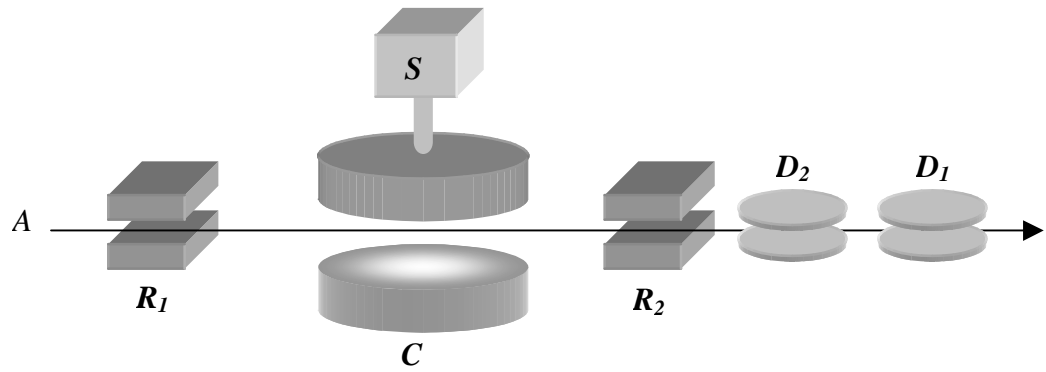
Fig.1. Sketch of the experimental setup for atom-driven field interaction.

Fig. 2a. Wigner function obtained when $\Theta=\pi$, $\alpha=0$, $\kappa
=\chi/20$, and $\chi\tau=2.06$. The variances for the quadrature operators
read $\left\langle \Delta X_{1}\right\rangle =\left\langle \Delta
X_{2}\right\rangle =2.63$ and the squeezing factor attained is $r(t_{2})=1.45
$. The mean photon number is $2.96$.

Fig. 2b. Wigner function obtained when $\Theta=\pi$, $\alpha=\sqrt{2}$,
$\kappa=\chi/20$, and $\chi\tau=3.99$.The variances for the quadrature
operators read $\left\langle \Delta X_{1}\right\rangle =5.58$ and
$\left\langle \Delta X_{2}\right\rangle =5.93$ and the squeezing factor is
$r(t_{2})=1.55$. The mean photon number is $26.38$.

Fig. 2c. Wigner function obtained when $\Theta=0$, $\alpha=5$, $\kappa
=\chi/20$, and $\chi\tau=1.68$. The variances for the quadrature operators
read $\left\langle \Delta X_{1}\right\rangle =0.31$ and $\left\langle \Delta
X_{2}^{2}\right\rangle =32.0$ and the squeezing factor is $r(t_{2})=1.5$. The
mean photon number is $256.17$.

Fig3a. Wigner function of the state generated by passing $N=2$ atoms through a
cavity initially prepared in the coherent state $\left|  \alpha\right|  =7.4 $
with $r=0.99$, leading to the maximized photon distribution function
$\mathcal{P}_{n=8}(t)=0.95$.

Fig3b. Wigner function of the state generated by passing $N=3$ atoms through a
cavity initially prepared in the coherent state $\left|  \alpha\right|  =8$
with $r=0.67$, leading to the maximized photon distribution function
$\mathcal{P}_{n=16}(t)=0.99$.

\newpage

\textbf{Tables }

Table I. The fidelity of the states $\left|  \Psi_{N=3}(t)\right\rangle $
generated from our model ($\mathcal{F}_{q}$) and those derived from the
standard circular states technique ($\emph{F}$), for different values of the
desired number states $\left|  q2^{3}\right\rangle $.

\begin{center}%
\begin{tabular}
[c]{cc}\hline\hline
$~q~$ & $~~~\mathcal{F}~~~~~\emph{F}~~~$\\\hline
\end{tabular}

\begin{tabular}
[c]{ccc}%
$1$ & $~0.99~$ & $~0.98~$\\
$3$ & $0.98$ & $0.65$\\
$5$ & $0.96$ & $0.50$\\
$10$ & $0.90$ & $0.35$\\\hline\hline
\end{tabular}

\end{center}


\begin{thebibliography}{99}                                                                                               %
\bibitem {Brune1}M. Brune, E. Hagley, J. Dreyer, X. Maitre, A. Maali, C.
Wunderlich, J. M. Raimond, and S. Haroche, Phys. Rev. Lett. \textbf{77}, 4887 (1996).

\bibitem {Monroe2}C. Monroe, D. M. Meekhof, B. E. King, and D. J. Wineland,
Science \textbf{272}, 1131 (1996).

\bibitem {Raimond}J. M. Raimond, M. Brune, S. Haroche, Phys. Rev. Lett.
\textbf{79}, 1964 (1997).

\bibitem {Brattke}S. Brattke, B. T. H. Varcoe, and H. Walther, Phys. Rev.
Lett. \textbf{86}, 3534 (2001).

\bibitem {Brune2}M. Brune, F. Schmidt-Kaler, A. Maali, J. Dreyer, E. Hagley,
J. M. Raimond, S. Haroche, Phys. Rev. Lett. \textbf{76}, 1800 (1996).

\bibitem {Knight1}P. Knight, Nature \textbf{380}, 392 (1996).

\bibitem {Monroe1}C. Monroe, D. M. Meekhof, B. E. King, W. M. Itano, and D. J.
Wineland, Phys. Rev. Lett. \textbf{75}, 4714 (1995).

\bibitem {Meekhof}D. M. Meekhof, C. Monroe, B. E. King, W. M. Itano, and D. J.
Wineland, Phys. Rev. Lett. \textbf{76}, 1796 (1996).

\bibitem {Leibfried}D. Leibfried, D. M. Meekhof, B. E. King, C. Monroe, W. M.
Itano, and D. J. Wineland, Phys. Rev. Lett. \textbf{77}, 4281 (1996).

\bibitem {Barnett}S. M. Barnett and D. T. Pegg, Phys. Rev. Lett. \textbf{76},
4148 (1996); B. Baseia, M. H. Y. Moussa, and V. S. Bagnato, Phys. Lett. A
\textbf{231}, 331 (1997).

\bibitem {Pegg}D. T. Pegg, L. S. Phillips, and S. M. Barnett, Phys. Rev. Lett.
\textbf{81}, 1604 (1998); C. J. Villas-B\^{o}as, Y. Guimar\~{a}es, M. H. Y.
Moussa, and B. Baseia, Phys. Rev. \textbf{A} 63, 055801 (2001).

\bibitem {D'Ariano}G. M. D'Ariano , L. Maccone, M. G. A. Paris, and M. F.
Sacchi, Fortschr. Phys. \textbf{48}, (2000).

\bibitem {Kwiat}P. G. Kwiat, K. Mattle, H. Weinfurter, A. Zeilinger, A. V.
Sergienko, and Y. Shih, Phys. Rev. Lett. \textbf{75}, 4337 (1995).

\bibitem {Dik}D. Bouwmeester, J. W. Pan, M. Daniell, H. Weinfurter, and A.
Zeilinger, Phys. Rev. Lett. \textbf{82}, 1345 (1999).

\bibitem {Teleportation}D. Bouwmeester, J.-W. Pan, K. Mattle, M. Eibl, H.
Weinfurter, and A. Zeilinger, Nature \textbf{390}, 575 (1997); D. Boschi, S.
Branca, F. De Martini, L. Hardy, and S. Popescu, Phys. Rev. Lett. \textbf{80},
1121 (1998); Y. H. Kim, S. P. Kulik, and Y. Shih, \textit{ibid.} \textbf{86},
1370 (2001); A. Furusawa, J. L. Sorensen, S. L. Braunstein, C. A. Fuchs, H. J.
Kimble, E. S. Polzik, Science \textbf{282}, 706 (1998).

\bibitem {Norton1}C. J. Villas-B\^{o}as, N. G. de Almeida, and M. H. Y.
Moussa, Phys. Rev. A \textbf{60}, 2759 (1999).

\bibitem {Norton2}R. M. Serra, C. J. Villas-B\^{o}as, N. G. de Almeida, and M.
H. Y. Moussa, J. Opt. B: Quantum Semiclass. Opt. \textbf{4}, 316 (2002).

\bibitem {Comm}J. I. Cirac, P. Zoller, H. J. Kimble, and H. Mabuchi, Phys.
Rev. Lett. \textbf{78}, 3221 (1997); T. Pellizzari, \textit{ibid}.
\textbf{79}, 5242 (1997).H. J. Briegel, W. D\"{u}r, J. I. Cirac, and P.
Zoller, \textit{ibid}. \textbf{81}, 5932 (1998); S. J. van Enk, H. J. Kimble,
J. I. Cirac, and P. Zoller, Phys. Rev. A \textbf{59}, 2659 (1999).

\bibitem {Comp}J. I. Cirac and P. Zoller, Phys. Rev. Lett. \textbf{74}, 4091
(1995); Q. A. Turchette, C. J. Hood, W. Lange, H. Mabuchi, and H. J. Kimble,
ibid. \textbf{75}, 4710 (1995); I. L. Chuang, L. M. K. Vandersypen, X. Zhou,
D. W. Leung, and S. Lloyd, Nature, \textbf{393}, 143 (1998); B. E. Kane,
\textit{ibid.} \textbf{393}, 143 (1998); L. M. K. Vandersypen, M. Steffen, G.
Breyta, C. S. Yannoni, M. H. Sherwood, I. L. Chuang, \textit{ibid.}
\textbf{414}, 883 (2001).

\bibitem {Zurek}W. H. Zurek, Phys. Today \textbf{44}, 36 (1991); W. H. Zurek
and J. P. Paz, Phys. Rev. Lett \textbf{72}, 2508 (1994).

\bibitem {Caldeira}A. O. Caldeira and A. J. Leggett, Annals of Physics
\textbf{149}, 374 (1983); A. O. Caldeira and A. J. Laggett, Physica A
\textbf{121}, 587 (1983).

\bibitem {POA}N. G. de Almeida, R. Napolitano, and M. H. Y. Moussa, Phys. Rev.
A \textbf{62}, 033815 (2000), N. G. de Almeida, P. B. Ramos, R. M. Serra, and
M. H. Y. Moussa, J. Opt. B: Quantum Semiclass. Opt. \textbf{2}, 792 (2000).

\bibitem {Bonifacio}R. Bonifacio, S. Olivares, P. Tombesi, D. Vitali, Phys.
Rev. A \textbf{61}, 053802 (2000).

\bibitem {Serra}R. M. Serra, P. B. Ramos, N. G. de Almeida, W. D. Jos\'{e},
and M. H. Y. Moussa, Phys. Rev. A \textbf{63}, 053813 (2001).

\bibitem {Poyatos}J. F. Poyatos, J. I. Cirac, and P. Zoller, Phys. Rev. Lett.
\textbf{77}, 4728 (1996); A. R. R. Carvalho, P. Milman, R. L. de Matos Filho,
and L. Davidovich, Phys. Rev. Lett. \textbf{86}, 4988 (2001).

\bibitem {Matos}A. R. R. Carvalho, P. Milman, R. L. de Matos Filho, and L.
Davidovich, Phys. Rev. Lett. \textbf{86}, 4988 (2001).

\bibitem {Lutkenhaus}N. Lutkenhaus, J. I. Cirac, and P. Zoller, Phys. Rev. A
\textbf{57}, 548 (1998).

\bibitem {Agarwal}G. S. Agarwal, M. O. Scully, and H. Walther, Phys. Rev.
Lett. \textbf{86}, 4271 (2001).

\bibitem {Myatt}C. J. Myatt, B. E. King, Q. A. Turchette, C. A. Sackett, D.
Kielpinski, W. M. Itano, C. Monroe, and D. Wineland, Nature \textbf{403}, 269 (2000).

\bibitem {JICirac}J. I. Cirac, Nature \textbf{413}, 375 (2001).

\bibitem {Zurek1}D. A. R. Dalvit, J. Dziarmaga, and W. H. Zurek, Phys. Rev. A
\textbf{62}, 013607 (2000)

\bibitem {Eberly}C. K. Law and J. H. Eberly, Phys. Rev. Lett., \textbf{76},
1055 (1996); F. L. Li and S.Y. Gao, Phys. Rev. A \textbf{62}, 043809 (2000).

\bibitem {Scully}M. O. Scully and S. Zubairy, \textit{Quantum Optics
(}Cambridge University Press, Cambridge, England, 1997); D. F. Walls and J.
Milburn, \textit{Quantum Optics} (Springer-Verlag, Berlin, 1994).

\bibitem {Dodonov}V. V. Dodonov, J. Opt. B: Quantum Semiclass. Opt.
\textbf{4,} R1 (2002).

\bibitem {Janszki}J. Janszky, P. Domokos, and P. Adam, Phys. Rev. A
\textbf{48}, 2213 (1993).

\bibitem {Parametric}C. J. Villas-B\^{o}as, N. G. de Almeida, R. M. Serra and
M. H. Y. Moussa, quant-ph/0303119.

\bibitem {98}R. E. Slusher, \textit{et al}., Phys. Rev. Lett. \textbf{55},
2409 (1985).

\bibitem {99}R. M. Shelby, \textit{et al}., Phys. Rev. Lett. \textbf{57}, 691 (1986).

\bibitem {100}L. A. Wu, \textit{et al}., Phys. Rev. Lett. \textbf{57}, 2520 (1986).

\bibitem {Meystre}P. Meystre and M. S. Zubairy, Phys. Lett. A \textbf{89}, 390
(1982); J. R. Kuklinski and J. L. Madajczyk, Phys. Rev. A \textbf{37}, R3175
(1988); M. Hillery, Phys. Rev. A \textbf{39}, 1556 (1989).

\bibitem {LDavid}L. Davidovich, A. Maali, M. Brune, J. M. Raimond, and S.
Haroche, Phys. Rev. Lett. \textbf{71}, 2360 (1993).

\bibitem {RMP}J. M. Raimond, M. Brune, and S. Haroche, Rev. Mod. Phys.
\textbf{73}, 565 (2001).

\bibitem {Holland}M. J. Holland, D. F. Walls, and P. Zoller, Phys. Rev. Lett.
\textbf{67}, 1716 (1994).

\bibitem {Walls}H. J. Carmichael, G. J. Milburn, and D. F. Walls, J. Phys. A:
Math Gen. \textbf{17}, 469 (1984).

\bibitem {Salomon}S. S. Mizrahi, M. H. Y. Moussa, and B. Baseia, Int. J. Mod.
Phys. B \textbf{8}, 1563 (1994).

\bibitem {Piza}A. F. R. de Toledo Piza, Phys. Rev. A \textbf{51}, 1612 (1995).

\bibitem {Aryeh}M. Zahler and Y. Ben Aryeh, Phys. Rev. A \textbf{43}, 6368 (1991).

\bibitem {Lewis}H. R. Lewis and W. B. Riesenfeld, J. Math. Phys. \textbf{10},
1458 (1969).

\bibitem {Puri}R. R. Puri and S. V. Lawande, Phys. Lett A \textbf{70}, 69 (1979).

\bibitem {Baseia}B. Baseia, M. H. Y. Moussa, and V. S. Bagnato, Phys. Lett. A
\textbf{231}, 331 (1997).

\bibitem {Piza1}Ji Il Kim, M. C. Nemes, A. F. R. de Toledo Piza, H. E. Borges,
Phys. Rev. Lett. \textbf{77}, 207 (1996).

\bibitem {Knight}M. S. Kim, W. Son, V. Buzek, and P. L. Knight, Phys. Rev. A
\textbf{65}, 032323 (2002).

\bibitem {CRM}C. J. Villas-B\^{o}as, R. M. Serra, and M. H. Y. Moussa, in preparation.

\bibitem {Kim}M. S. Kim and V. Buzek, Phys. Rev. A \textbf{47}, 610 (1993).

\bibitem {Brune3}M. Brune \textit{et al}., Phys. Rev. A \textbf{45}, 5193 (1992).

\bibitem {Vitali}P. Tombesi and D. Vitali, Phys. Rev. A \textbf{50}, 4253 (1994).

\bibitem {Carugno}G. Carugno (private communication).
\end{thebibliography}
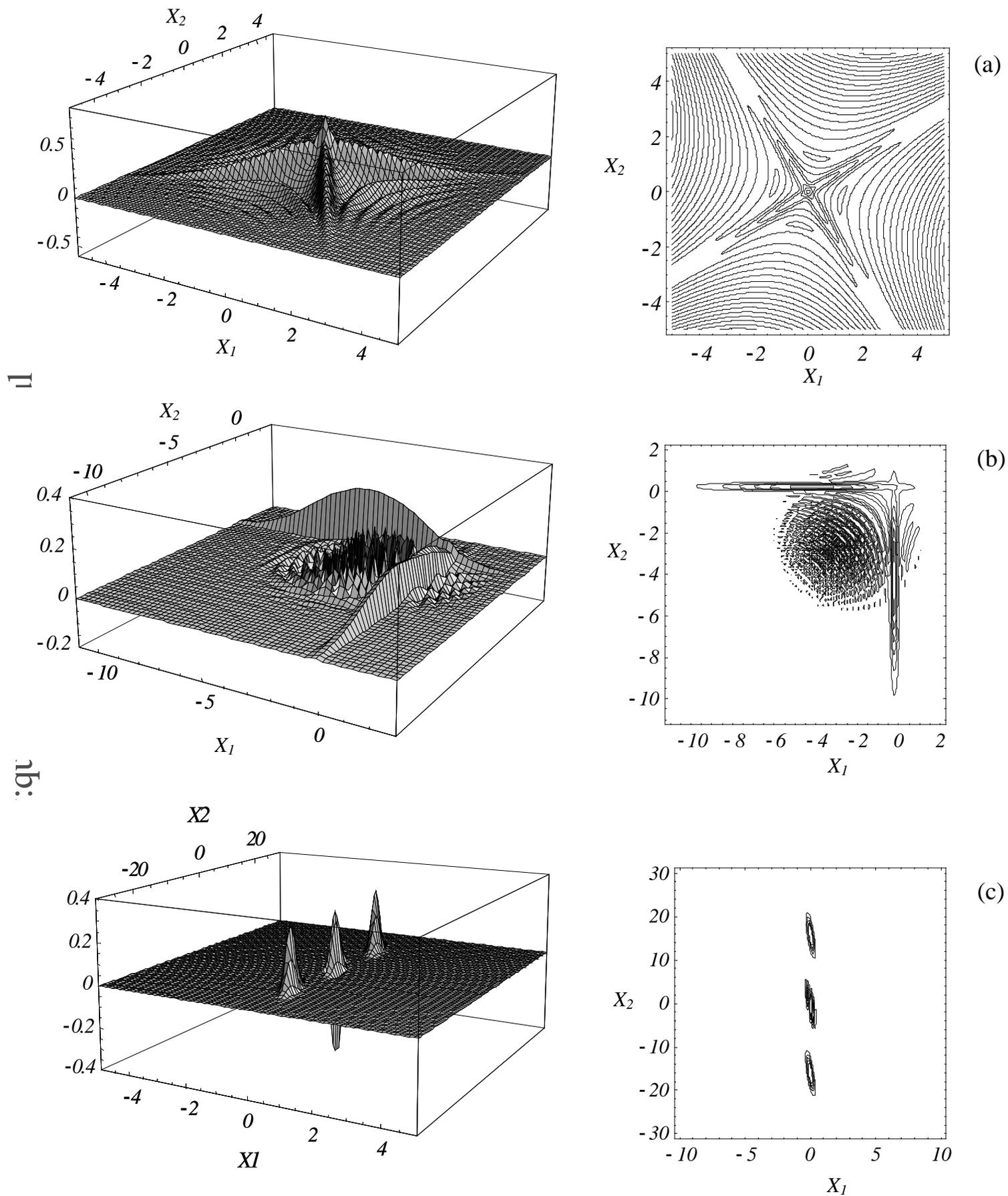
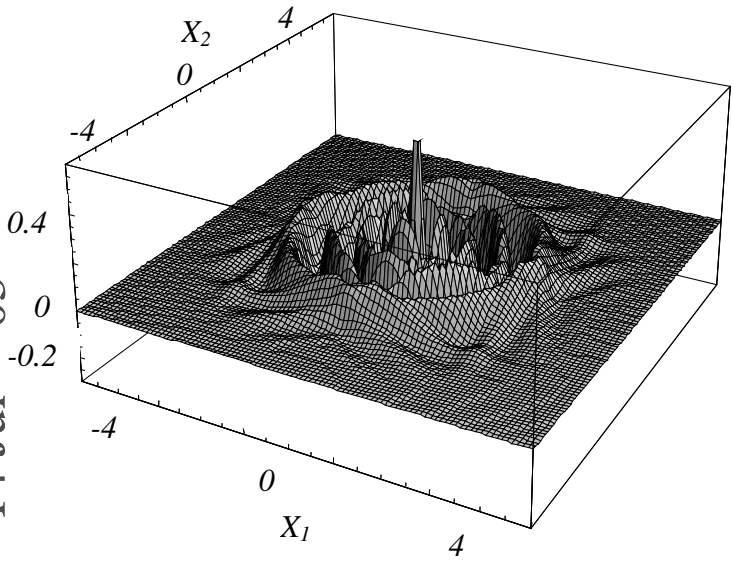
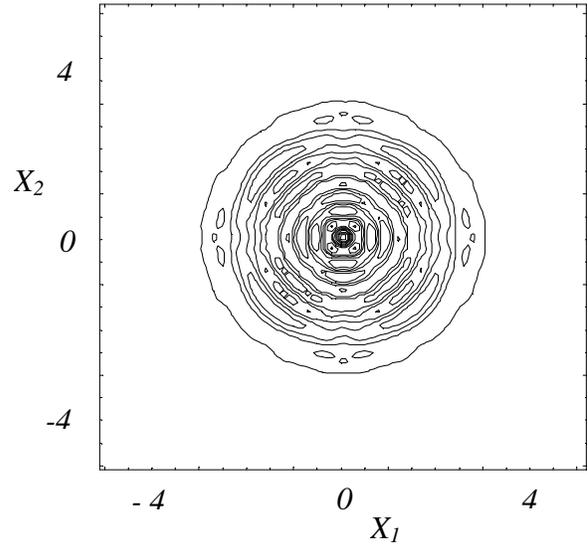
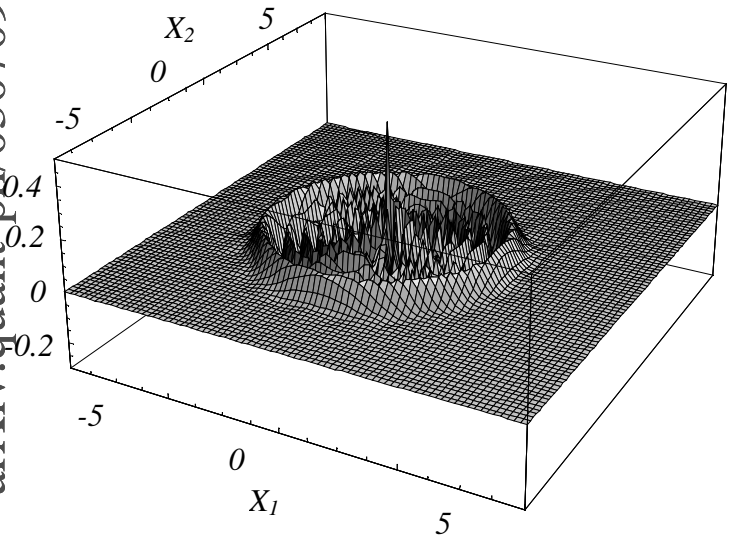
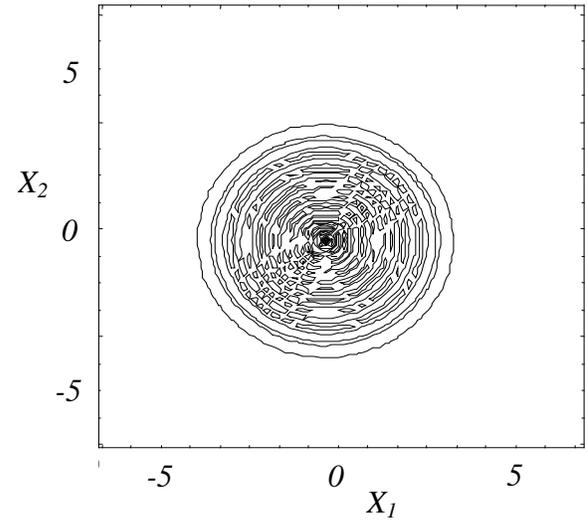
\end{document}